\documentclass[aps,reprint,amsmath,amssymb,nofootinbib,longbibliography]{revtex4-2}

\usepackage{bm,color}
\usepackage{graphicx}
\usepackage{microtype}
\usepackage{tikz}
\usetikzlibrary{arrows.meta,positioning,calc}

\makeatletter
\let\revtex@normalsize\normalsize
\def\normalsize{%
  \revtex@normalsize
  \abovedisplayskip      5pt \@plus 1pt \@minus 2pt
  \belowdisplayskip      5pt \@plus 1pt \@minus 2pt
  \abovedisplayshortskip 2pt \@plus 1pt \@minus 1pt
  \belowdisplayshortskip 3pt \@plus 1pt \@minus 1pt
}%
\normalsize
\def\section{\@startsection{section}{1}{\z@}%
  {0.32cm \@plus 1ex \@minus .2ex}{0.22cm}{\normalfont\small\bfseries}}%
\def\subsection{\@startsection{subsection}{2}{\z@}%
  {0.28cm \@plus 1ex \@minus .2ex}{0.16cm}{\normalfont\small\bfseries}}%
\def\subsubsection{\@startsection{subsubsection}{3}{\z@}%
  {0.35cm \@plus 1ex \@minus .2ex}{0.20cm}{\normalfont\small\itshape}}%
\makeatother

\providecommand{\arraybackslash}{\let\\\tabularnewline}
\newcommand{\rgcell}{\raggedright\arraybackslash}

\def\_#1{{\bf #1}}
\newcommand{\dyad}[1]{\overline{\overline{#1}}}
\let\h\^
\def\^#1{\hat{\bf #1}}
\def\.{\cdot}
\def\x{\times}
\def\d{\partial}
\def\o{\omega}
\def\E{\varepsilon}
\def\M{\mu}
\def\D{\nabla}
\def\e{\begin{equation}}
\def\f{\end{equation}}
\def\l#1{\label{eq:#1}}
\def\r#1{(\ref{eq:#1})}
\def\Re{\mathop{\rm Re}}
\def\Im{\mathop{\rm Im}}
\newcommand{\avg}[1]{\left\langle #1\right\rangle}
\newcommand{\ii}{\mathrm i}
\newcommand{\cL}{\mathcal L}
\newcommand{\cA}{\mathcal A}
\newcommand{\cP}{\mathcal P}
\newcommand{\cS}{\mathcal S}
\newcommand{\cH}{\mathcal H}

\newcommand{\Ie}{{\rm e}}

\newcommand{\Ia}{{\rm a}}

\newcommand{\SMref}{the Supplemental Material~\cite{SuppMat}}

\begin{document}

\title{Dispersive Corrections to Electromagnetic Energy and Power Flux\\
  Are Imprints of Eliminated Degrees of Freedom}

\author{Stanislav Maslovski}
\affiliation{Department of Electronics, Telecommunications and Informatics and Instituto de Telecomunica\c{c}\~oes,\\ University of Aveiro, 3810-193 Aveiro, Portugal}

\author{Constantin Simovski}
\affiliation{Department of Electronics and Nanoengineering, School of Electrical Engineering,\\ Aalto University, FI-00076 Espoo, Finland}

\date{\today}

\begin{abstract}
  In a spatially dispersive medium the electromagnetic energy and flux are given by the
  frequency and wave-vector derivatives of the constitutive parameters, a result known as
  the Agranovich--Ginzburg (AG) formulas. Here, we clarify the physical meaning of the
  derivative terms in these relations and show that they measure the energy and flux
  carried by the degrees of freedom of the internal field-matter dynamics that the
  Maxwellian constitutive description hides. We construct an extended state-space model of
  such dynamics for a wide class of temporally and spatially dispersive media, from which
  the Maxwellian constitutive relations follow upon elimination of the internal variables,
  the effective material operator resulting from the Schur complement of the eliminated
  internal block. A congruence identity relates this complement to the extended state-space
  description, which allows us to extend the AG formulas to general bianisotropic media and
  to establish the conditions of their validity in the presence of dissipation. The AG
  dispersive corrections are thus the imprints of the eliminated
  degrees of freedom rather than independent constitutive postulates. They hold for fields
  driven at arbitrary real frequency and wave vector, and not only for eigenmodes, provided
  that the sources act on the fields alone and that any dissipation is carried by the
  electromagnetic subsystem. We
  verify the construction on the Lorentz, Drude, magnetized-plasma, and hydrodynamic
  models, the last of which describes a nonlocal metal. Additionally, a bianisotropic
  $\Omega$-medium and a capacitively loaded wire medium are treated in the Supplemental Material.
\end{abstract}

\maketitle

\section{Introduction}
\label{sec:intro}

As is well known, the energy density of a time-harmonic electromagnetic field in a
dispersive medium is not $(\E|\_E|^2+\M|\_H|^2)/4$. Brillouin's formula replaces $\E$ in
this expression by $\d(\o\E)/\d\o$~\cite{Brillouin1960,LL8}. In spatially dispersive
media, as was established on phenomenological grounds by Agranovich and
Ginzburg~\cite{Agranovich1966}, the energy density keeps this form with a
wave-vector-dependent $\dyad\E(\o,\_k)$, whereas the power flux acquires, in addition to
the Poynting term, a contribution proportional to the wave-vector derivative of the
effective permittivity. We refer to this pair of relations as the Agranovich--Ginzburg
(AG) formulas. The physics behind the derivative terms in these relations has nevertheless
been examined only in particular models, such as the Lorentz
medium~\cite{Loudon1970,Ruppin2002} or spatially dispersive
crystals~\cite{Bishop1976,Forstmann1979,Nelson1996}. Although the phenomenological route by which these formulas
were obtained hints at {\em how} the dependence of the constitutive parameters on $\o$
and $\_k$ should enter the energy balance, it says nothing about {\em what} the derivative
corrections comprise in general.

A related question, of long standing, is which expressions for the energy density, the
power flux and the heating rate are the correct ones. The Poynting theorem relates the
three without determining them: the flux is fixed only up to a solenoidal field, and the
balance by itself does not divide the remaining terms between storage and dissipation.
Different macroscopic separations have accordingly been proposed, and have been shown to
lead to different physical conclusions~\cite{Markel2008,Markel2010,Richter2010,SilveirinhaReply2010}.

A first-principles derivation of the macroscopic Poynting vector, heating rate, and stored
energy was given in Ref.~\cite{Silveirinha2009} for composite media formed by dielectric
and metallic inclusions, covering cases of weak and strong spatial dispersion including
artificial magnetism and bianisotropy. Starting from the microscopic Maxwell equations in a
periodic structured material and defining the macroscopic fields by an averaging procedure
equivalent to extraction of the lowest-order Floquet--Bloch harmonic, that work established
a mathematically exact relation between quadratic expressions of the microscopic
fields---such as the cell-averaged microscopic Poynting vector---and the macroscopic fields
together with the effective dielectric function, thereby placing the AG expressions on a
microscopic footing and showing that they survive even for strong spatial
dispersion~\cite{Belov2003,Maslovski2009} rather than a small correction.

Two questions were left open in that study (and were only partially covered in
Refs.~\cite{Silveirinha2012,Silveirinha2021}, which we discuss below).
The first, and the one that motivates the present work, is what the derivative corrections
{\em are}: a derivation of this kind confirms them but does not explain their nature, in
that it does not identify which physical quantities $\d(\o\dyad\E)/\d\o$ and
$\d\dyad\E/\d\_k$ measure. The second concerns
the scope of the construction, which rests on a Floquet--Bloch decomposition and hence on
strict periodicity at the microscopic level, whereas strongly nonlocal media such as dense quasi-periodic or
amorphous wire arrays possess well-defined macroscopic fields and the known quasistatic
homogenization models~\cite{Maslovski2009} can be expected to apply to them without
requiring a strictly periodic lattice.

\begin{figure*}[t]
\includegraphics[width=0.9\textwidth]{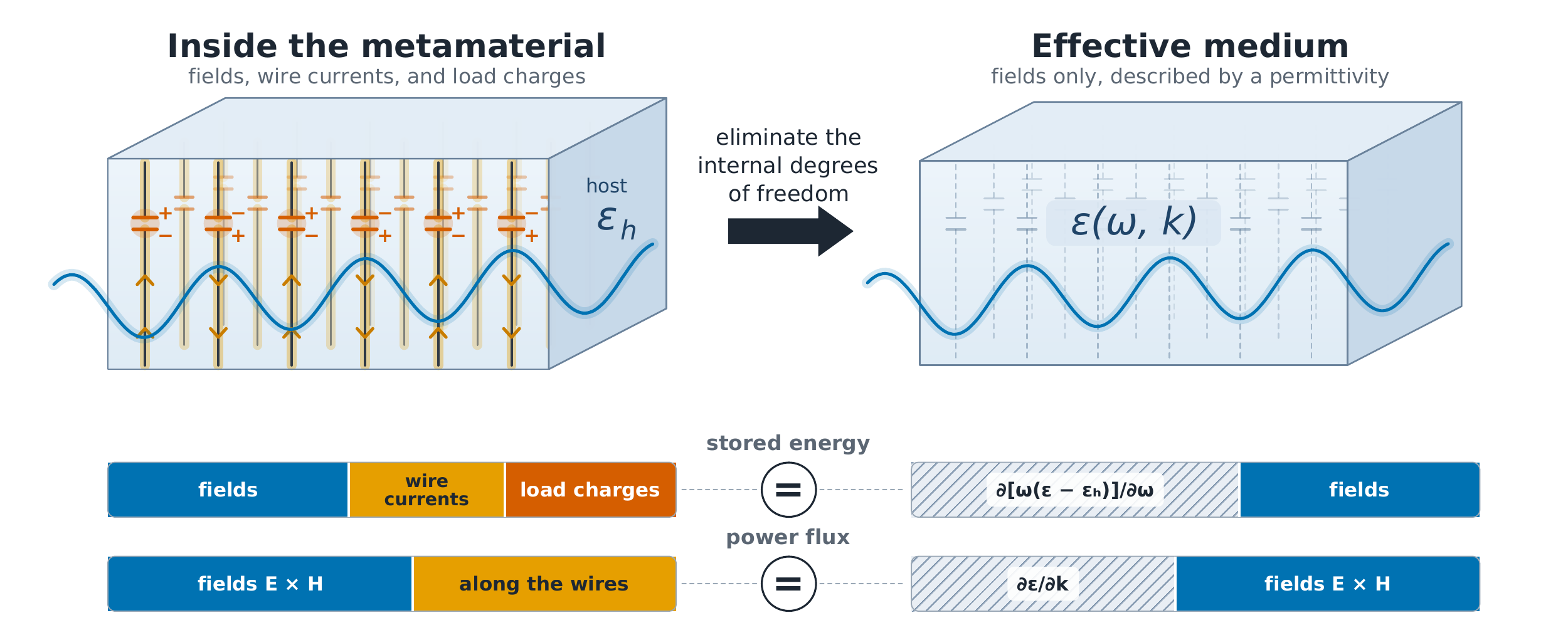}
\caption{Where the energy and power flux of a wave reside before and after the internal
  degrees of freedom of a metamaterial are eliminated. Left: in a capacitively loaded wire
  medium embedded in a host of permittivity $\E_h$, a wave stores energy in the fields, in
  the currents and potentials of the wires, and in the charges on the loads, and carries
  power both in the fields
  and along the wires. Right: the effective-medium description retains only the fields and
  the permittivity $\E(\o,\_k)$. The energy and power flux of the eliminated variables
  reappear, exactly, as the dispersive terms of the Agranovich--Ginzburg formulas, which
  involve $\d[\o(\E-\E_h)]/\d\o$ and $\d\E/\d\_k$. This medium is worked out in
  \SMref.}
\label{fig:concept}
\end{figure*}

In this paper we show that both questions have a common answer for a broad class of media.
If the macroscopic dynamics closes as a local, first-order system in an extended state
space, with a positive-definite energy metric, then the energy density, the power flux,
and the dissipation rate are quadratic forms of the extended state vector whose matrices
are independent of frequency and wave vector. The familiar derivative corrections appear
only when the additional state variables are eliminated in favor of an effective
Maxwellian material operator, and they then carry exactly the energy and power flux of the
eliminated variables; Fig.~\ref{fig:concept} illustrates this for the wire medium of
\SMref. The groundwork for these ideas was laid in earlier works that constructed effective Hamiltonian or Lagrangian descriptions in extended state space to describe the effects of {\em temporal} dispersion~\cite{Tip1998,Figotin2005,Figotin2007,Raman2010,Philbin2010,Silveirinha2021} (discussed in more detail below). This work treats both temporal and spatial dispersion equally within a unified, self-consistent formalism. We make the above argument precise: the Maxwellian operator is the Schur
complement~\cite{Haynsworth1968a,Zhang2005} of the matrix pencil in the extended state
space, taken with respect to the internal (auxiliary) state variables. Consequently, a
congruence relates this complement to the pencil of the extended problem. In a medium with
loss, its anti-Hermitian part gives the heating rate, whereas its derivatives with respect
to frequency and wave vector give the AG energy density and power flux whenever no loss is
associated with the eliminated degrees of freedom. The equality of the energy transport and group
velocities~\cite{Welters2014,Yaghjian2018} is independent of all this: it holds already in the extended state space,
before any elimination.

Hamiltonian extensions of {\em temporally} dispersive and dissipative media, in which
auxiliary fields render the dynamics conservative, were developed by Tip and by Figotin
and Schenker~\cite{Tip1998,Figotin2005,Figotin2007}; from such a formulation the latter
obtained closed-form energy and flux densities in terms of the auxiliary fields, together
with an approximate Brillouin-type expression for a quasi-monochromatic wave. Auxiliary
fields of a similar kind had been used to recast the band-structure problem of a temporally
dispersive metamaterial as a Hermitian eigenvalue problem~\cite{Raman2010}. A Lagrangian
in which a continuum of auxiliary oscillators accounts for temporal dispersion and
dissipation was constructed by Philbin~\cite{Philbin2010}, who also obtained, for a
lossless medium and without auxiliary fields, an exact energy and momentum balance from
which the Brillouin expression is recovered in the monochromatic limit~\cite{Philbin2011}. Working instead
from passivity and the analytic properties of the response functions, Welters, Avniel and
Johnson established a generalized Brillouin formula and the equality of the energy
transport and group velocities for bianisotropic media with a local
response~\cite{Welters2014}; their argument uses the ordinary Poynting flux and, as they
note, does not extend to a general nonlocal response. For lossless continua with spatial
dispersion, Yaghjian showed that the AG flux, the one containing the wave-vector derivative
of the permittivity, defines an energy transport velocity that again equals the group
velocity~\cite{Yaghjian2018}.

An extended-state formulation of lossless temporally dispersive media that shares some similarity
to ours in its formal apparatus was developed by Silveirinha~\cite{Silveirinha2021},
although its premises are totally different. It starts from the frequency-domain material
matrix, written in a partial-fraction expansion form with a finite number of poles, and
introduces auxiliary variables that reduce the temporal dynamics of the material response
to a system of first-order equations; the electromagnetic field and these variables
collected into an extended state vector then obey a dynamic matrix equation with a
constant energy metric, from which a Poynting theorem for the extended state follows. An
energy density expressed through the frequency derivative of the material matrix is
recovered from it by pairing two fields oscillating at slightly different frequencies and
letting the difference vanish, a divided-difference counterpart of the Brillouin formula;
the same framework gives the canonical momentum of the field in a medium and modal
expansions of the Green function, which lie outside the scope of the present work. In the
terminology introduced below, the spatial operator of that formulation acts on the
electromagnetic block alone, the auxiliary variables couple to the fields but not to one
another, and the sources act on the fields alone; spatial dispersion is not developed
there, although an extension to a subclass of spatially dispersive materials is noted.

An early precursor to the present work was Ref.~\cite{Silveirinha2012}, in which the
excitation of a uniaxial wire medium by elementary sources was considered. That work
employed a local state-space model with the wire current and an additional potential as
internal degrees of freedom. It was shown that such a formalism naturally leads to an exact
Poynting theorem in which the part of the energy density stored in the wires and the part of
the power flux carried along them are expressed through these variables. Upon elimination of the auxiliary variables, the cycle average
of the power flux was stated, without presenting a derivation, to reduce to the AG expression for
lossless spatially dispersive media. The model itself, and the local state-space
description of wire media on which the example given in \SMref\ is built, were introduced
in Refs.~\cite{Maslovski2009,Maslovski2010}.

The present work generalizes these results to the whole class of media with temporal and
spatial dispersion and dissipation, proves them, and explains them. The construction runs in the direction
opposite to that of Ref.~\cite{Silveirinha2021}: rather than assuming a form for the
material operator and building auxiliary variables that reproduce it, we start from the
local first-order system and obtain the material operator from it, as a Schur complement,
so that its pole structure is a property of the internal block and not an input. The work
differs from those cited above in three further aspects. First, the identities obtained
below are exact and are stated for the Maxwellian material operator itself: they hold off
the dispersion surface, for fields driven at any real frequency and wave vector, and they
{\em admit loss of arbitrary strength} provided that it involves only the degrees of
freedom associated with the electromagnetic field, the remaining case being controlled by
an error term given in closed form. Second, general {\em spatial} dispersion is treated on
the same footing as temporal dispersion, for the whole class of media at once, with
internal variables that may transport energy and couple to one another, which is what
produces the wave-vector derivative of the AG flux. The extended-state formulations cited
above were developed for temporal
dispersion~\cite{Tip1998,Figotin2005,Figotin2007,Raman2010,Philbin2010,Silveirinha2021};
the spatially dispersive AG flux has been established for lossless
continua~\cite{Yaghjian2018}; and internal variables carrying part of the power flux have
been treated before only in particular models, namely the polarization waves of spatially
dispersive crystals~\cite{Bishop1976,Forstmann1979,Nelson1996} and the wire
medium~\cite{Silveirinha2012}.
Third, and most important for the question raised above, the mechanism is identified: the
elimination of the internal state by a Schur complement is what converts a fixed metric
into the frequency and wave-vector derivative expressions of the AG formulas, and this is
what those derivatives measure---the energy and the power flux carried by the eliminated
degrees of freedom.

Viewed in this way, the AG corrections are an instance of a pattern that is not specific
to electromagnetism. Whenever a subset of the degrees of freedom of a linear problem in Fourier space is
eliminated in favor of the remaining ones, the reduced description acquires a dependence
on the spectral parameters, and quantities that were simple and dispersionless in the full
problem acquire derivatives with respect to these parameters in the reduced one.
The Schur complement is the algebraic form of that elimination; it appears as Foster's
reactance theorem for a one-port electrical network~\cite{Foster1924}, in which the stored
energy is proportional to the frequency derivative of the driving-point reactance, as the Feshbach
projection in scattering and many-body theory~\cite{Feshbach1958,Feshbach1962}, and as the
self-energy in quasiparticle theory~\cite{FetterWalecka,Hedin1965}, where the spectral weight
of the quasiparticle pole is likewise determined by a frequency derivative. When the self-energy depends on momentum as well, the
renormalized quasiparticle velocity acquires a contribution from its momentum derivative,
the counterpart of the wave-vector derivative that spatial dispersion adds to the AG power
flux.

The results are presented in order of decreasing generality, and the organization of the
paper follows that order. Section~\ref{sec:class} defines the class of media considered here and derives
an exact local and instantaneous conservation law that holds for arbitrary fields in an
inhomogeneous and dissipative medium. Section~\ref{sec:elimination} specializes to plane
waves in a uniform medium, driven at real but otherwise unconstrained $\o$ and $\_k$,
performs the elimination of the internal degrees of freedom, and illustrates it on four standard media with
their state-space matrices written out. Section~\ref{sec:dissipation} proves the
congruence identity, which is not limited by Hermitian symmetry and therefore holds with
dissipation, and draws from it the heating rate; the attenuation balance for a freely
decaying wave follows from the conservation law. Section~\ref{sec:imprints} narrows to a medium in which no loss is
associated with the internal variables, and derives
the derivative identity and with it the AG energy density and power flux, still
off the dispersion surface. Section~\ref{sec:velocity} reaches the narrowest case, a
source-free lossless medium, for which it proves that the energy transport velocity of an
eigenmode equals the group velocity, and carries the energy and flux identities from a single
plane wave over to the integral characteristics of a free field. Section~\ref{sec:hydroex} illustrates these results on the
hydrodynamic model of a nonlocal metal with collisional damping and verifies each of them
numerically, in the same order. A capacitively loaded wire medium is worked out in \SMref,
where the same sequence of checks is repeated for a metamaterial with lumped loads. Section~\ref{sec:scope}
discusses the range of validity of the results, what the material operator does and does not
determine about the energy and the power flux, the implications for homogenized models, and
possible extensions. Appendices~\ref{app:adjoint}--\ref{app:modal} contain the
intermediate algebra, and Appendix~\ref{app:hydronum} collects the parameters of the
numerical calculations.

Throughout, the time convention is $e^{\ii(\_k\.\_r-\o t)}$, so that $\d/\d t\to-\ii\o$
and $\d/\d x_j\to\ii k_j$ for a plane wave.

\section{Media with local state-space realization}
\label{sec:class}

\subsection{Governing system}
\label{sec:system}

Let the macroscopic dynamics (including electrodynamics) of a medium be defined by the following operator equation
\e
\hat{\cL}\_\Psi=-\mathsf M\frac{\d\_\Psi}{\d t}-\mathsf G\_\Psi+\_F_{\rm ext},
\qquad
\_\Psi=\begin{pmatrix}\_\Psi_\Ie\\ \_\Psi_\Ia\end{pmatrix},
\l{model}
\f
where the block-column 6-vector $\_\Psi_\Ie$ represents the state of the
electromagnetic field and $\_\Psi_\Ia$ collects the internal (auxiliary) state variables of the
medium in which the electromagnetic field propagates, with $\_\Psi$ being the extended
state vector. $\mathsf M$ and $\mathsf G$ are the constitutive matrices of the medium. As
shown below, $\mathsf M$ is associated with the energy storage and $\mathsf G$ with
dissipation. The vector $\_F_{\rm ext}=(\_F_\Ie,\,\_F_\Ia)^{\mathsf T}$ represents
possible external forces that drive the system.

We require that

(i) $\mathsf M$ and $\mathsf G$ be real symmetric and independent of time, with $\mathsf M$
positive definite and $\mathsf G$ positive semidefinite;

(ii) the operator $\hat{\cL}=\sum_j\mathsf L_j\,\d/\d x_j+\mathsf L_0$, which contains only
first-order derivatives with respect to the Cartesian coordinates $x_{j}$, $j=1,2,3$, and
has constant matrices $\mathsf L_j$, be formally skew-adjoint; since $\d/\d x_j$ is itself
formally skew-adjoint, this is equivalent to
\e
\mathsf L_j^{\mathsf T}=\mathsf L_j,
\qquad
\mathsf L_0^{\mathsf T}=-\mathsf L_0;
\l{symmetry}
\f
we write $\mathsf L_{1},\mathsf L_{2},\mathsf L_{3}$ and $\mathsf L_x,\mathsf L_y,\mathsf L_z$
interchangeably;

(iii) the realization of the internal dynamics be minimal.\footnote{Such a minimal system
  includes only the degrees of freedom that are controllable and observable through
  coupling to the electromagnetic fields. See also a related remark at the end of this
  subsection.}

The above restrictions on $\hat{\cL}$ alone are not enough to ensure that the
first two block rows of Eq.~\r{model} take the form of Maxwell's equations. For that, the
electromagnetic block of $\hat{\cL}$ must be such that a coordinate transformation brings
it to the 6-curl form,
\e
\big(\hat{\cL}-\mathsf L_0\big)\big|_{\Ie\Ie}
=\begin{pmatrix}0&\D\x\\ -\D\x&0\end{pmatrix} .
\l{curlblock}
\f
In what follows we assume that such a transformation has been made, so that the first two
block rows of Eq.~\r{model} are Faraday's and Amp\`ere's laws with $\_\Psi_\Ie$ being the
6-vector of physical electromagnetic fields, $\_\Psi_\Ie\equiv(\_H,\_E)^{\mathsf T}$.
This choice of the field variables is not unique. With $\_E$ and $\_B$ left untouched,
the substitution $\_H=\_H'+\d\_T/\d t$, $\_D=\_D'+\D\x\_T$ with an arbitrary differentiable
vector field $\_T$ preserves the form of Maxwell's equations. What it changes is the
constitutive relations and whatever follows from them, including the expressions for the
energy density and the Poynting vector; an example is given in Sec.~S1\,D of \SMref.

The equivalence between the formal skew-adjointness of $\hat{\cL}$ and the symmetry
conditions~\r{symmetry} is elementary and is verified in
Appendix~\ref{app:adjoint}. Note that although the $\mathsf L_{j}$ are constant
matrices, $\mathsf M$, $\mathsf G$, and $\mathsf L_0$ may depend on position, so that
inhomogeneous media are included; only the plane-wave analysis below requires them to be
uniform.

Equation~\r{model} is a Friedrichs symmetric hyperbolic system with a dissipative
zeroth-order term~\cite{Friedrichs1954}, and it can also be viewed as a linear port-Hamiltonian system
with a quadratic storage function and dissipation rate~\cite{vanDerSchaft2006}. Conditions
(i) and (ii) are not merely convenient. As shown in Sec.~\ref{sec:balance}, they entail a
local and instantaneous energy balance with a positive stored energy and a non-negative
dissipation rate, both obtained from the governing equations of wave motion rather than
postulated. We shall refer to Eq.~\r{model} as a {\em local state-space realization} of the
medium, and to the media that admit one satisfying (i)--(iii) as the media of the class
considered here. The word {\em local} refers to the realization and not to the medium: a
medium of this class is in general strongly nonlocal when described in terms of the
electromagnetic fields alone, and relating the two descriptions is the object of what
follows.

Condition (iii) is important because for a minimal realization without loss the stored
energy expression is unique~\cite{Willems1972a,Willems1972b,Foster1924}. Since Eq.~\r{model} is a model of the medium, the energy density it assigns might be
expected to depend on how that model was set up. It does not. Any two minimal realizations of the
same {\em externally controlled} behavior are related by a change of state-space coordinates, and the storage
function---here $\frac12\_\Psi^{\mathsf T}\mathsf M\_\Psi$ derived in the next subsection---is invariant under it. For a
uniform medium these statements apply, at each real $\_k$, to the finite-dimensional system
obtained by Fourier transformation in space, and minimality is understood in that sense.
Hence
the storage function $\frac12\_\Psi^{\mathsf T}\mathsf M\_\Psi$, and with it the energy
density below, is fixed by the medium itself rather than by the particular lossless minimal
realization chosen for it; what is invariant is this quadratic functional and not the
individual entries of $\mathsf M$, which change with the state-space coordinates. A non-minimal realization would
append internal modes that are either invisible from the fields or undriven by them;
these carry energy of their own without affecting the electromagnetic response, and the
uniqueness is lost precisely because such modes may be added at will.

As will be shown in Sec.~\ref{sec:imprints}, a unique correspondence between the energy
and power flux of the full state-space description and those of the reduced Maxwellian
description requires the external forces $\_F_\Ia$ acting directly on the internal degrees
of freedom to vanish. When $\_F_\Ia\ne0$, part of the internal state is set by these
sources rather than by the electromagnetic fields, and no expression built from the fields
and the Maxwellian constitutive relations alone can account for the energy it carries.
This further delimits the applicability domain of the generalized AG formulas derived in
Sec.~\ref{sec:imprints}.

The same requirement excludes static internal states on which $\_F_\Ie$ alone may not act,
like pre-charged capacitors in the capacitively loaded wire medium of \SMref.
Indeed, in the Laplace-transform formulation, a nonzero initial state of a dynamic variable
enters the transformed equations as a source in the internal block, that is, as a
contribution to $\_F_\Ia$, which is excluded by the above requirement.

\subsection{Poynting theorem in extended state space}
\label{sec:balance}

Contracting Eq.~\r{model} with $\_\Psi^{\mathsf T}$ and using Eq.~\r{symmetry} gives an
exact local and instantaneous conservation law. The $\mathsf L_0$ contribution drops out
because that matrix is antisymmetric: since $\_\Psi^{\mathsf T}\mathsf L_0\_\Psi$ is a scalar,
it equals its own transpose, so that
\e
\_\Psi^{\mathsf T}\mathsf L_0\_\Psi
=\_\Psi^{\mathsf T}\mathsf L_0^{\mathsf T}\_\Psi
=-\_\Psi^{\mathsf T}\mathsf L_0\_\Psi
\ \Longrightarrow\
\_\Psi^{\mathsf T}\mathsf L_0\_\Psi=0 .
\l{L0vanishes}
\f
This is the algebraic reason why the couplings between the fields and the internal
variables, which reside in $\mathsf L_0$, exchange power without producing or destroying it.
The first-order derivative terms resulting from the left-hand side of Eq.~\r{model} are exact divergences, because $\mathsf L_j$ is symmetric and
constant,
\e
\frac{\d}{\d x_j}\left(\_\Psi^{\mathsf T}\mathsf L_j\_\Psi\right)
=2\,\_\Psi^{\mathsf T}\mathsf L_j\frac{\d\_\Psi}{\d x_j} ,
\l{divergence}
\f
so that $\_\Psi^{\mathsf T}\hat{\cL}\_\Psi=\D\.\_S$. On the right-hand side of
Eq.~\r{model}, the symmetry and time independence of $\mathsf M$ give
$\_\Psi^{\mathsf T}\mathsf M\,\d_t\_\Psi=\d_t(\frac12\_\Psi^{\mathsf T}\mathsf M\_\Psi)$.
Collecting the terms,
\e
\D\.\_S=-\frac{\d W}{\d t}-Q+P_{\rm ext} ,
\l{Poyntingtheorem}
\f
with
\begin{align}
W&=\tfrac12\_\Psi^{\mathsf T}\mathsf M\_\Psi,
\quad
S_j=\tfrac12\_\Psi^{\mathsf T}\mathsf L_j\_\Psi,
\l{quadratic1}\\
Q&=\_\Psi^{\mathsf T}\mathsf G\_\Psi,
\quad
P_{\rm ext} = \_\Psi^{\mathsf T}\_F_{\rm ext}.
\l{quadratic2}
\end{align}
Here $\_S$ is the power flux density (a generalized Poynting vector), $W$ is the energy density, $Q$ is the dissipation
(heating) rate, and $P_{\rm ext}$ is the power of external sources. The flux $\_S$ contains
the ordinary Poynting vector $\_E\x\_H$, which comes from the curl entries of $\hat{\cL}|_{\Ie\Ie}$, and,
whenever $\mathsf L_j$ couples internal variables to one another, the power carried by those
variables, such as the power transported along the wires in the wire-medium example of
\SMref. The above expressions
for these instantaneous quantities are exact, hold for arbitrary fields, and contain no
derivatives of any material parameter. In the extended state space the constitutive
relations are local and instantaneous, and no separate contributions associated with
temporal or spatial dispersion appear.

No property of $\mathsf M$, $\mathsf G$ or $\mathsf L_j$ beyond Eq.~\r{symmetry} and the
symmetry of $\mathsf M$ has been used; in particular, positive definiteness of $\mathsf M$
is needed only to interpret $W$ as an energy, and positive semidefiniteness of $\mathsf G$
only to interpret $Q$ as a dissipation rate. Since $\mathsf M$, $\mathsf G$ and $\mathsf L_0$
may depend on position, Eq.~\r{Poyntingtheorem} holds for inhomogeneous media as well. The
constancy of $\mathsf L_j$ is a simplifying assumption. Were $\mathsf L_j$ position dependent, one
would have $\_\Psi^{\mathsf T}\mathsf L_j\d_j\_\Psi=\d_jS_j-\frac12\_\Psi^{\mathsf T}(\d_j\mathsf
L_j)\_\Psi$, and the spatial operator would have to be taken in the symmetrized form
$\sum_j\frac12(\mathsf L_j\d_j+\d_j\mathsf L_j)+\mathsf L_0$ to remain skew-adjoint and to restore
Eq.~\r{Poyntingtheorem}; we do not pursue this generalization since it is of little
relevance for the present work.

For time-harmonic processes, since $\mathsf M$, $\mathsf L_j$ and $\mathsf G$ are real,
symmetric, and independent of time and frequency, the cycle averages of
Eqs.~\r{quadratic1} and~\r{quadratic2} are the same quadratic forms multiplied by $1/2$
(Appendix~\ref{app:adjoint}),
\e
\avg{W}=\tfrac14\_\Psi^{\dagger}\mathsf M\_\Psi ,
\quad
\avg{S_j}=\tfrac14\_\Psi^{\dagger}\mathsf L_j\_\Psi ,
\quad
\avg{Q}=\tfrac12\_\Psi^{\dagger}\mathsf G\_\Psi ,
\l{avgs}
\f
the different numerical prefactor of $\avg{Q}$ originating solely from the absence of the
factor $\frac12$ in the definition of $Q$ in Eq.~\r{quadratic2}.

Let us note that the same symbol $\_\Psi$ is used in the time and frequency
domains. The distinction between the quadratic forms written for the two domains is carried
by the Hermitian conjugation: a transpose $\_\Psi^{\mathsf T}$ always indicates an
expression involving the real instantaneous state, as in Eq.~\r{quadratic1} and in
Sec.~S2\,B of \SMref, and a Hermitian conjugate $\_\Psi^{\dagger}$ always indicates a
complex amplitude expression, as in Eq.~\r{avgs}. What makes the two sets of expressions
identical up to the prefactors is that $\mathsf M$, $\mathsf L_j$ and $\mathsf G$ are real,
symmetric, and independent of time and hence of frequency; no derivative of a material
parameter can therefore enter at this stage. Although a complex amplitude may in general
carry a spatial dependence, in the sections that deal with plane waves in a uniform medium
$\_\Psi$ denotes the constant vector multiplying $e^{\ii(\_k\.\_r-\o t)}$. For real $\_k$
the two forms of expressions for $\_\Psi$ differ by a phase factor that cancels from every quadratic form; for
complex $\_k$ an amplitude factor appears that does not cancel, and this is taken up in
Sec.~\ref{sec:attenuation}.

\section{Elimination of internal variables}
\label{sec:elimination}

\subsection{Driven plane-wave problem}
\label{sec:pencil}

Let the medium now be uniform, with $\mathsf G\neq0$ in general, and let it be driven by a
distributed source $\_F_{\rm ext}\propto e^{\ii(\_k\.\_r-\o t)}$, where $\_k$ and $\o$ are
real-valued. In such a source-driven problem, $\_k$ and $\o$ are not required to satisfy any
dispersion relation and are independent of each other. With $\d/\d t \to -\ii\o$ and
$\d/\d x_j \to \ii k_j$, Eq.~\r{model} becomes
\e
\left(\ii\sum_j k_j\mathsf L_j+\mathsf L_0\right)\_\Psi
=\ii\o\mathsf M\_\Psi-\mathsf G\_\Psi+\_F_{\rm ext} ,
\f
and, multiplying by $-\ii$ and collecting the terms,
\e
  \hat{\cP}(\o,\_k)\,\_\Psi = \tilde{\_F}_{\rm ext} ,
  \l{Peq}
\f
where the matrix pencil $\hat{\cP}(\o,\_k)$ is defined as
\begin{align}
  \hat{\cP}(\o,\_k) &= \hat{\cA}(\_k)-\o\mathsf M-\ii\mathsf G ,
  \l{Pdef}\\
  \hat{\cA}(\_k) &= \sum_j k_j\mathsf L_j-\ii\mathsf L_0 .
\end{align}
In Eq.~\r{Peq} the constant factor $-\ii$ has been absorbed into the renormalized spectral
amplitude of the source $\tilde{\_F}_{\rm ext}$; no result depends on this renormalization,
which merely fixes the phase reference of the source. In what follows we therefore drop the
tilde above $\_F_{\rm ext}$.

We partition the pencil in accordance with how we partition the state,
\e
\hat{\cP}=
\begin{pmatrix}
\hat{\cP}_{\Ie\Ie} & \hat{\cP}_{\Ie\Ia}\\
\hat{\cP}_{\Ia\Ie} & \hat{\cP}_{\Ia\Ia}
\end{pmatrix}.
\l{Pblocks}
\f
For real $\_k$ the matrix $\hat{\cA}$ is Hermitian: the $\mathsf L_j$ are real and symmetric,
hence Hermitian, whereas $(-\ii\mathsf L_0)^{\dagger}=\ii\mathsf L_0^{\mathsf T}=-\ii\mathsf L_0$ by
Eq.~\r{symmetry}. Consequently the anti-Hermitian part of the pencil is carried entirely by
the dissipation matrix,
\e
\frac{\hat{\cP}-\hat{\cP}^{\dagger}}{2\ii}=-\mathsf G ,
\l{antiP}
\f
a relation used repeatedly below. In the lossless case $\mathsf G=0$ the pencil is
Hermitian.

\begin{table*}[tb]
\caption{\label{tab:examples}Local state-space realizations of four standard dispersive
  media and of the loaded wire medium of \SMref, each with its dissipation. In every row the
  state is
  $\_\Psi=(\_H,\_E,\_\Psi_\Ia)^{\mathsf T}$, the electromagnetic block of $\mathsf M$ is
  ${\rm diag}(\M_0\dyad I,\E_0\dyad I)$ [${\rm diag}(\M_0\dyad I,\E_h\dyad I)$ for the wire
  medium], and the curl entries of $\mathsf L_j$, common to all media, are omitted. The third
  column gives the internal block $\mathsf M_{\Ia\Ia}$. In the last column
  $\mathsf X_{(Y,Z)}$ denotes the block of the matrix $\mathsf X$ in the row of variable $Y$
  and the column of variable $Z$; the transposed block is not listed, being
  $\mathsf X_{(Z,Y)}=-[\mathsf X_{(Y,Z)}]^{\mathsf T}$ for $\mathsf L_0$ and
  $+[\mathsf X_{(Y,Z)}]^{\mathsf T}$ for the symmetric $\mathsf L_j$ and $\mathsf G$, by
  Eq.~\r{symmetry}. Setting $\mathsf G=0$ gives the lossless realization in every case.
  Vector-valued
  blocks such as $\^z/A$ or $\^x_j$ couple a vector variable to a scalar one. Notation:
  $\dyad I$ is the $3\x3$ identity, $\o_0$ and $\o_p$ are the resonance and plasma
  frequencies, $[\_\Omega\x]$ is the matrix of the cross product with the cyclotron vector
  $\_\Omega=e\_B_0/m$ ($e>0$ the elementary charge), $\nu$ is a collision rate, $\beta$ is
  the hydrodynamic velocity, $R$ the resistance per unit length of a wire,
  $\^x_j$ are the Cartesian unit vectors, and the wire current $I$, the additional wire
  potential $\varphi$, the load charge $q_\ell$ and the array parameters $L$, $C$, $A$ and
  $\gamma_\ell$ are defined in Secs.~S2\,A and~S2\,B of \SMref. Each realization satisfies Eq.~\r{symmetry} and returns the standard
  permittivity upon the elimination procedure of Sec.~\ref{sec:schur}.}
\begin{ruledtabular}
\begin{tabular}{llp{0.19\textwidth}p{0.44\textwidth}}
Medium & $\_\Psi_\Ia$ & \rgcell $\mathsf M_{\Ia\Ia}$ & \rgcell other nonzero blocks of $\mathsf L_0$, $\mathsf L_j$ and $\mathsf G$\\
\colrule
Lorentz & $(\_P,\_J)$, $\_J=\d_t\_P$
 & \rgcell ${\rm diag}\!\left(\dfrac{\o_0^2}{\E_0\o_p^2}\dyad I,\ \dfrac{1}{\E_0\o_p^2}\dyad I\right)$
 & \rgcell $(\mathsf L_0)_{(\_E,\_J)}=+\dyad I$, $(\mathsf L_0)_{(\_J,\_P)}=\dfrac{\o_0^2}{\E_0\o_p^2}\dyad I$, $\mathsf G_{(\_J,\_J)}=\dfrac{\nu}{\E_0\o_p^2}\dyad I$\\[2mm]
Drude & $\_J$
 & \rgcell $\dfrac{1}{\E_0\o_p^2}\dyad I$
 & \rgcell $(\mathsf L_0)_{(\_E,\_J)}=+\dyad I$, $\mathsf G_{(\_J,\_J)}=\dfrac{\nu}{\E_0\o_p^2}\dyad I$\\[2mm]
Magnetized plasma & $\_J$
 & \rgcell $\dfrac{1}{\E_0\o_p^2}\dyad I$
 & \rgcell $(\mathsf L_0)_{(\_E,\_J)}=+\dyad I$, $(\mathsf L_0)_{(\_J,\_J)}=-\dfrac{[\_\Omega\x]}{\E_0\o_p^2}$, $\mathsf G_{(\_J,\_J)}=\dfrac{\nu}{\E_0\o_p^2}\dyad I$\\[2mm]
Hydrodynamic metal & $(\_J,\varrho)$
 & \rgcell ${\rm diag}\!\left(\dfrac{1}{\E_0\o_p^2}\dyad I,\ \dfrac{\beta^2}{\E_0\o_p^2}\right)$
 & \rgcell $(\mathsf L_0)_{(\_E,\_J)}=+\dyad I$, $(\mathsf L_j)_{(\_J,\varrho)}=\dfrac{\beta^2}{\E_0\o_p^2}\^x_j$, $\mathsf G_{(\_J,\_J)}=\dfrac{\nu}{\E_0\o_p^2}\dyad I$\\[2mm]
Loaded wire medium & $(I,\varphi,q_\ell)$
 & \rgcell ${\rm diag}\!\left(\dfrac{L}{A},\ \dfrac{C}{A},\ \gamma_\ell\right)$
 & \rgcell $(\mathsf L_0)_{(\_E,I)}=\^z/A$, $(\mathsf L_0)_{(I,q_\ell)}=\gamma_\ell$, $(\mathsf L_3)_{(I,\varphi)}=1/A$, $\mathsf G_{(I,I)}=R/A$\\
\end{tabular}
\end{ruledtabular}
\end{table*}

\subsection{Schur complement and material operator}
\label{sec:schur}

Let us now eliminate the internal state variables to obtain the usual Maxwellian
description. Writing the source as $\_F_{\rm ext}=(\_F_\Ie,\_F_\Ia)^{\mathsf T}$ and
assuming $\hat{\cP}_{\Ia\Ia}$ to be invertible, the lower block row of
$\hat{\cP}\_\Psi=\_F_{\rm ext}$ gives
\e
\_\Psi_\Ia=\hat{\cP}_{\Ia\Ia}^{-1}\_F_\Ia
-\hat{\cP}_{\Ia\Ia}^{-1}\hat{\cP}_{\Ia\Ie}\_\Psi_\Ie ,
\l{auxstate}
\f
and substitution into the upper block row leaves
\e
\hat{\cS}\_\Psi_\Ie=\_F_\Ie-\hat{\cP}_{\Ie\Ia}\hat{\cP}_{\Ia\Ia}^{-1}\_F_\Ia ,
\quad
\hat{\cS}=\hat{\cP}_{\Ie\Ie}
-\hat{\cP}_{\Ie\Ia}\hat{\cP}_{\Ia\Ia}^{-1}\hat{\cP}_{\Ia\Ie} ,
\l{schur}
\f
where $\hat{\cS}$ is the Schur complement of $\hat{\cP}_{\Ia\Ia}$ in the pencil
$\hat{\cP}$. Note that Eq.~\r{auxstate} does not require $\_\Psi_\Ie$ to be an
eigenvector: it expresses the internal variables in terms of the fields and the sources for
{\em any} $\_\Psi_\Ie$. That is why the results below are valid for driven
fields with arbitrary real $\o$ and $\_k$.

To identify $\hat{\cS}$, note that, by the choice of coordinates made in
Sec.~\ref{sec:system}, Eq.~\r{curlblock}, the first-order part of $\hat{\cA}$ reproduces
the curl operators in the electromagnetic block. Writing $(\D\x)\to\ii\dyad K$ with
$\dyad K\.\_a=\_k\x\_a$, we have
\e
\sum_j k_j\mathsf L_j\bigg|_{\Ie\Ie}
=\begin{pmatrix}0&\dyad K\\ -\dyad K&0\end{pmatrix} ,
\f
which is Hermitian because $\dyad K$ is real and antisymmetric,
$\dyad K^{\dagger}=-\dyad K$. All remaining contributions to $\hat{\cS}$, those of
$\mathsf M$ and $\mathsf G$ and those that arise from the elimination, are collected, by
definition, into the material operator
\e
\dyad{\cal M}(\o,\_k)\equiv
-\frac{1}{\o}\left[
\hat{\cS}-\begin{pmatrix}0&\dyad K\\ -\dyad K&0\end{pmatrix}
\right],
\quad
\dyad{\cal M}=
\begin{pmatrix}\dyad\M&\dyad\zeta\\ \dyad\xi&\dyad\E\end{pmatrix},
\l{Mcaldef}
\f
so that
\e
\hat{\cS}=-\o\dyad{\cal M}(\o,\_k)+\begin{pmatrix}0&\dyad K\\ -\dyad K&0\end{pmatrix} .
\l{Sofcal}
\f
The physical relevance of this definition follows from writing out the source-free
form $\hat{\cS}\_\Psi_\Ie=0$ explicitly:
\e
\dyad K\.\_E-\o\left(\dyad\M\.\_H+\dyad\zeta\.\_E\right)=0
\ \Longleftrightarrow\
\_k\x\_E=\o\_B ,
\f
\e
-\dyad K\.\_H-\o\left(\dyad\xi\.\_H+\dyad\E\.\_E\right)=0
\ \Longleftrightarrow\
\_k\x\_H=-\o\_D ,
\f
which are precisely Maxwell's equations for a plane wave in a bianisotropic medium with
$\_B=\dyad\M\.\_H+\dyad\zeta\.\_E$ and $\_D=\dyad\E\.\_E+\dyad\xi\.\_H$. The medium is in
general bianisotropic and spatially dispersive. In the lossless case $\hat{\cS}$ is
Hermitian, and, since the $\dyad K$ block is Hermitian by itself, $\dyad{\cal M}$ is
Hermitian as well; this requires $\dyad\E$ and $\dyad\M$ to be Hermitian and
$\dyad\xi=\dyad\zeta^{\dagger}$, which is the standard losslessness condition for
bianisotropic media.

The elimination requires $\hat{\cP}_{\Ia\Ia}$ to be invertible. Where it is not, the internal
block has a nontrivial kernel at that $(\o,\_k)$, and whether this reflects a property of the
medium or of the model is settled by condition~(iii). In a minimal realization every internal
degree of freedom is observable through its coupling to the electromagnetic fields, so that
$\hat{\cP}_{\Ie\Ia}$ does not annihilate that kernel and the Schur complement~\r{schur}
diverges: the singularity is a constitutive pole, at which the material operator ceases to
exist. In a non-minimal realization this need not follow, since an internal variable
uncoupled from the fields contributes a kernel that $\hat{\cP}_{\Ie\Ia}$ does annihilate,
leaving $\dyad{\cal M}$ finite and unaltered. A constitutive pole is therefore a failure of
the reduced description and not of Eq.~\r{model}, which remains regular there; what this
implies for the results below is taken up in Sec.~\ref{sec:validity}. The case $\o=0$, which
definition~\r{Mcaldef} excludes in any case, needs a separate remark. In the lossless Drude,
magnetized-plasma and hydrodynamic models of Table~\ref{tab:examples} it is an ordinary
constitutive pole, the familiar $-\o_p^2/\o^2$ divergence of the permittivity. In the
loaded wire medium, by contrast, the kernel of the internal block at $\o=0$ consists of the
static internal states that Sec.~\ref{sec:system} excludes together with $\_F_\Ia$.

\subsection{Standard dispersive media in state-space form}
\label{sec:examples}

The class of media that admit Eq.~\r{model} is broad. To show this, we proceed here in the
direction opposite to that of the previous subsection: we start from familiar Maxwellian
models and construct their extended state-space realizations. Table~\ref{tab:examples} collects the
internal variables and the matrices of four standard media, together with the loaded wire
medium treated in \SMref, each with its dissipation. In every case the internal variables are quantities
with an immediate physical meaning, the entries of $\mathsf M$ and $\mathsf G$ reproduce the
known stored energies and loss mechanisms, and the elimination procedure of Sec.~\ref{sec:schur} returns
the known effective permittivity. The normalization of each internal variable is not free:
multiplying its equation of motion by a constant does not change the physics, and requiring
Eq.~\r{symmetry} fixes that constant, as shown for the hydrodynamic model in
Appendix~\ref{app:hydro} and for the wire medium in Sec.~S2\,B of \SMref.

A Lorentz medium belongs to the class with $\_\Psi_\Ia=(\_P,\_J)^{\mathsf T}$, where $\_P$
is the medium polarization and $\_J=\d\_P/\d t$ the polarization current, the oscillator
equation $\d\_J/\d t=-\o_0^2\_P+\E_0\o_p^2\_E$ being first order in this pair. The
corresponding entries of $\mathsf M$ give
$(\o_0^2|\_P|^2+|\_J|^2)/(2\E_0\o_p^2)$, the familiar potential and kinetic energies of the
bound oscillators. The Drude metal is the limit $\o_0\to0$, in which $\_P$ leaves the state
altogether and only the current remains, with
$|\_J|^2/(2\E_0\o_p^2)=m|\_J|^2/(2ne^2)$ the kinetic energy of the free carriers of mass
$m$, charge $-e$ and density $n$. In both, a collision rate $\nu$ damps the current alone,
giving the Joule loss $Q=\nu|\_J|^2/(\E_0\o_p^2)$.

The case of a magnetized plasma~\cite{Stix1992} follows from the Drude case by adding the cyclotron term
$\_\Omega\x\_J$, $\_\Omega=e\_B_0/m$, to the current equation. Since the matrix of a cross
product is antisymmetric, this term belongs to $\mathsf L_0$ and not to $\mathsf M$: the static
magnetic field organizes the exchange of power among the internal variables but stores no
energy, as it must, since a static magnetic field does no work. Gyrotropy, and with it the
antisymmetric part of the Maxwellian permittivity tensor, thus originates entirely in the
zeroth-order block.

A magnetized ferrite is described in the same way, with the transverse magnetization as the
internal variable: the linearized Landau--Lifshitz equation~\cite{Gurevich1996} is of the
form~\r{model}, with the precession frequency $\gamma\M_0H_0$, set by the gyromagnetic ratio
$\gamma$ and the static internal field $H_0$, in the place of the cyclotron frequency, and
the elimination of the magnetization returns the Polder permeability
tensor~\cite{Polder1949}, so that ferromagnetic resonance and magnetic gyrotropy are
accommodated exactly as the cyclotron resonance and the electric gyrotropy of a magnetized
plasma are, with the dispersive correction appearing in the permeability rather than in the
permittivity. Writing the ferrite in the form~\r{model} requires a preliminary step that the
examples above do not---Faraday's law has to be recombined with the Landau--Lifshitz
equation for condition~(i) to hold---and we reserve a detailed discussion, together with
that of spin waves, for a future work.

The media considered so far are dispersive in time but remain spatially local even in the
reduced description. The hydrodynamic model of a nonlocal metal~\cite{Raza2015} is the
first example in which the spatial derivatives of $\hat{\cL}$ act on an internal
variable, which produces the wave-vector dependence of the effective material response. The
linearized electron fluid obeys
\begin{subequations}
\l{hydro}
\begin{align}
\frac{\d\_J}{\d t}&=\E_0\o_p^2\_E-\beta^2\D\varrho-\nu\_J ,
\l{hydroJ}\\
\frac{\d\varrho}{\d t}&=-\D\.\_J ,
\l{hydrorho}
\end{align}
\end{subequations}
where $\_J$ is the induced current density, $\varrho$ the induced charge density,
$\beta$ the hydrodynamic velocity and $\nu$ the collision rate of the electron fluid, the
lossless medium being recovered at $\nu=0$. The first two are related to the polarization
$\_P$ of the electron fluid,
\e
\_J=\frac{\d\_P}{\d t},
\qquad
\varrho=-\D\.\_P ,
\l{hydroP}
\f
so that, for a time-harmonic field, $\_P=\_J/(-\ii\o)$. Written for $\_P$ alone, the model
reads $\d^2\_P/\d t^2+\nu\,\d\_P/\d t=\E_0\o_p^2\_E+\beta^2\D(\D\.\_P)$, which is of second order in both
time and space and does not fit Eq.~\r{model}. The two internal variables remove one order
each: $\_J$ replaces the second time derivative, and $\varrho$ the second spatial
derivative, so that the system becomes of first order in both, which is what
conditions (i) and (ii) ask for. Eliminating them, on the other hand, leaves a relation
between $\_D$ and $\_E$ alone that is a convolution in time as well as in space; the
internal variables thus restore temporal locality together with spatial locality. The
resulting entries of
$\mathsf M$, listed in Table~\ref{tab:examples}, are the kinetic energy of the electron
fluid and its compressional energy $\beta^2\varrho^2/(2\E_0\o_p^2)=m\beta^2(\delta n)^2/(2n)$,
where $n$ and $\delta n$ are the equilibrium and perturbed electron densities
(Appendix~\ref{app:hydro}), and
the elimination returns the standard permittivity dyadic of a nonlocal electron fluid,
uniaxial about the direction of the wave vector,
\e
\dyad\E(\o,\_k)=\E_{\rm T}(\o)\left(\dyad I-\^k\^k\right)
+\E_{\rm L}(\o,k)\,\^k\^k ,
\quad
\^k=\frac{\_k}{k} ,
\l{hydroeps}
\f
with the transverse and longitudinal permittivities
\e
\begin{split}
\E_{\rm T}&=\E_0\left(1-\frac{\o_p^2}{\o^2+\ii\nu\o}\right),\\
\E_{\rm L}&=\E_0\left(1-\frac{\o_p^2}{\o^2+\ii\nu\o-\beta^2k^2}\right),
\end{split}
\l{hydroepsLT}
\f
or, in a single expression,
\e
\begin{split}
\dyad\E(\o,\_k)=\E_0\Bigg[\dyad I-\o_p^2\Bigg(&
\frac{\dyad I-\^k\^k}{\o^2+\ii\nu\o}\\
&+\frac{\^k\^k}{\o^2+\ii\nu\o-\beta^2k^2}\Bigg)\Bigg].
\end{split}
\l{hydroepsone}
\f
The entire wave-vector dependence resides in the longitudinal part, since $\E_{\rm T}$
coincides with the permittivity of a Drude metal and does not depend on $\_k$. The derivative
$\d\dyad\E/\d\_k$ therefore arises only from $\^k\^k$ and from $\E_{\rm L}$. This example is
worked out in detail in Sec.~\ref{sec:hydroex}.
The loaded wire medium of \SMref\ is a fifth example, in which the internal
variables are the averaged wire current, the additional wire potential, and the charge
stored in the loads.

More generally, the class contains any linear medium whose response can be decomposed into
internal state variables governed by partial differential equations of first order in both
time and space, provided that conditions (i)--(iii) of Sec.~\ref{sec:system} are satisfied.

\section{Heating rate}
\label{sec:dissipation}

\subsection{Congruence identity}
\label{sec:congruence}

From here to the end of Sec.~\ref{sec:imprints} we take $\_F_\Ia=0$, that is, we assume
the medium to be driven through the electromagnetic sources and fields alone. Equation~\r{auxstate} then
reduces to $\_\Psi_\Ia=-\hat{\cP}_{\Ia\Ia}^{-1}\hat{\cP}_{\Ia\Ie}\_\Psi_\Ie$, so that
the whole state is determined by its electromagnetic part. Introducing the rectangular
matrix
\e
\hat T=
\begin{pmatrix}
\dyad I\\
-\hat{\cP}_{\Ia\Ia}^{-1}\hat{\cP}_{\Ia\Ie}
\end{pmatrix},
\qquad\text{so that}\qquad
\_\Psi=\hat T\_\Psi_\Ie ,
\l{Tdef}
\f
and writing $\hat T_\Ia\equiv-\hat{\cP}_{\Ia\Ia}^{-1}\hat{\cP}_{\Ia\Ie}$, we expand
\e
\hat T^{\dagger}\hat{\cP}\hat T
=\hat{\cP}_{\Ie\Ie}
+\hat{\cP}_{\Ie\Ia}\hat T_\Ia
+\hat T_\Ia^{\dagger}\hat{\cP}_{\Ia\Ie}
+\hat T_\Ia^{\dagger}\hat{\cP}_{\Ia\Ia}\hat T_\Ia .
\l{expand}
\f
The second term equals
$-\hat{\cP}_{\Ie\Ia}\hat{\cP}_{\Ia\Ia}^{-1}\hat{\cP}_{\Ia\Ie}$, which is the Schur
correction we want. The last two terms cancel identically: factoring out
$\hat T_\Ia^{\dagger}$ on the left, we have
\e
\hat T_\Ia^{\dagger}\hat{\cP}_{\Ia\Ie}
+\hat T_\Ia^{\dagger}\hat{\cP}_{\Ia\Ia}\hat T_\Ia
=\hat T_\Ia^{\dagger}
\left[\hat{\cP}_{\Ia\Ie}
-\hat{\cP}_{\Ia\Ia}\hat{\cP}_{\Ia\Ia}^{-1}\hat{\cP}_{\Ia\Ie}\right]=0 ,
\l{cancel}
\f
since the bracket evidently vanishes. Therefore, the Schur complement $\hat{\cS}$, Eq.~\r{schur}, can be expressed as
\e
\hat{\cS}=\hat T^{\dagger}\hat{\cP}\hat T ,
\l{congruence}
\f
which is the congruence identity central to this work.

The structure of Eq.~\r{cancel} is what makes the congruence useful in the dissipative
case. The cancellation uses only the definition of
$\hat T_\Ia$ and the invertibility of $\hat{\cP}_{\Ia\Ia}$; the left factor
$\hat T_\Ia^{\dagger}$ is never evaluated. Consequently Eq.~\r{congruence} is an algebraic
identity valid for an {\em arbitrary} $\hat{\cP}$: no symmetry is required of
$\hat{\cP}_{\Ia\Ia}$, and none of the relation between $\hat{\cP}_{\Ie\Ia}$ and
$\hat{\cP}_{\Ia\Ie}$. Dissipation inside the medium, $\mathsf G_{\Ia\Ia}\ne0$, and
dissipative coupling between the fields and the internal variables,
$\mathsf G_{\Ie\Ia}\ne0$, are therefore both permitted.

\subsection{Heating rate in driven medium}
\label{sec:heat}

The derivation of the heating rate presented next uses only the congruence~\r{congruence}
and, therefore, holds for a medium with arbitrary dissipation. As before, we assume that the
medium is driven by a distributed plane-wave source with real $\o$ and $\_k$. Applying
$\hat T^{\dagger}(\cdot)\hat T$ to both sides of Eq.~\r{antiP} and using
Eq.~\r{congruence},
\e
\frac{\hat{\cS}-\hat{\cS}^{\dagger}}{2\ii}
=-\hat T^{\dagger}\mathsf G\hat T .
\l{antihermS}
\f
On the other hand, the $\dyad K$ block of Eq.~\r{Sofcal} is Hermitian, so that
\e
\frac{\hat{\cS}-\hat{\cS}^{\dagger}}{2\ii}
=-\o\,\frac{\dyad{\cal M}-\dyad{\cal M}^{\dagger}}{2\ii} .
\l{antihermM}
\f
Comparing the two expressions and contracting with $\_\Psi_\Ie$, we express the
electromagnetic heating rate in terms of the material operator,
\e
\avg{Q}=\frac12\_\Psi^{\dagger}\mathsf G\_\Psi
=\frac12\_\Psi_\Ie^{\dagger}\hat T^{\dagger}\mathsf G\hat T\_\Psi_\Ie
=\frac{\o}{2}\_\Psi_\Ie^{\dagger}
\frac{\dyad{\cal M}-\dyad{\cal M}^{\dagger}}{2\ii}\_\Psi_\Ie .
\l{AGheat}
\f
The sign in front of the right-hand side is fixed by the time convention, and it is
consistent with passivity. Indeed, combining Eq.~\r{antihermS} with Eq.~\r{antihermM},
\e
\o\,\frac{\dyad{\cal M}-\dyad{\cal M}^{\dagger}}{2\ii}
=\hat T^{\dagger}\mathsf G\hat T ,
\l{passivity}
\f
whose right-hand side is positive semidefinite because $\mathsf G$ is. For $\o>0$ the
anti-Hermitian part of the material operator is therefore a positive semidefinite matrix,
and Eq.~\r{AGheat} gives $\avg{Q}\ge0$ for {\em any} field, as passivity requires. For a
medium that is neither magnetic nor bianisotropic this reduces to the familiar statement
that the anti-Hermitian part of $\dyad\E$ alone is positive semidefinite. Had the opposite
convention $e^{+\ii\o t}$ been adopted, the right-hand sides of Eqs.~\r{antihermS} and
\r{AGheat} would acquire the opposite sign, and passivity would correspond to a negative
semidefinite anti-Hermitian part. Note that the passivity of the effective description is here a {\em
  consequence} of the positive semidefiniteness of $\mathsf G$ in the state-space model,
transported through the congruence, and not an independent assumption about
$\dyad{\cal M}$.

\subsection{Freely attenuated modes}
\label{sec:attenuation}

The restriction to real $\_k$ in Eq.~\r{antiP} is essential, since a free wave in a dissipative medium is attenuated. For $\_k=\_k'+\ii\_k''$ the matrix
$\hat{\cA}$ is no longer Hermitian, and Eq.~\r{Pdef} gives instead
\e
\frac{\hat{\cP}-\hat{\cP}^{\dagger}}{2\ii}
=\sum_j k_j''\mathsf L_j-\mathsf G ,
\l{antihermk}
\f
so that Eq.~\r{antihermS} acquires an extra term and Eq.~\r{AGheat} does not apply as it
stands. The physical content is nevertheless transparent: contracting Eq.~\r{antihermk} with
$\_\Psi$ and using
$\_\Psi^{\dagger}\hat{\cP}\_\Psi=\_\Psi^{\dagger}\hat{\cP}^{\dagger}\_\Psi=0$ on the
dispersion surface, the left-hand side vanishes and Eqs.~\r{avgs} give
\e
\avg{Q}=2\,\_k''\!\cdot\!\avg{\_S} ,
\l{attenuation}
\f
where $\avg{Q}$ and $\avg{\_S}$ now both carry the factor $e^{-2\_k''\.\_r}$ of the
attenuated wave, so that Eq.~\r{attenuation} holds pointwise. It is the balance between
dissipation and the spatial decay of the flux for a freely propagating attenuated wave. It
is the cycle average of the conservation law~\r{Poyntingtheorem} for such a wave: with $\avg{W}$ independent of time, $P_{\rm ext}=0$, and
$\D\.\avg{\_S}=-2\_k''\.\avg{\_S}$, Eq.~\r{Poyntingtheorem} reduces to
Eq.~\r{attenuation}. Like the conservation law, it is a statement in the extended state space and
holds for any loss mechanism.

Equation~\r{AGheat} describes a different situation: the driven problem at real $\o$ and
$\_k$, in which the dissipated power is supplied by the sources rather than by an
attenuating flux. This is also the setting of the classical treatments, in which the
dissipation rate of a time-harmonic field is written as a quadratic form of the field
amplitudes with the anti-Hermitian part of the constitutive tensor taken at real
frequency~\cite{LL8} and, for a spatially dispersive medium, at real wave vector as
well~\cite{Agranovich1966}. For weak dissipation the two cases are related perturbatively:
evaluating both $\avg{\_S}$, by Eq.~\r{AGflux}, and $\avg{Q}$, by Eq.~\r{AGheat}, with the
fields of the lossless mode, Eq.~\r{attenuation} yields the attenuation constant to first
order in $\mathsf G$ from the material operator alone.

\section{Energy density and power flux}
\label{sec:imprints}

\subsection{Derivative identity and its range of validity}
\label{sec:derivid}

Let $\lambda$ denote either $\o$ or one of the $k_j$. Differentiating Eq.~\r{congruence},
\e
\frac{\d\hat{\cS}}{\d\lambda}
=\left(\frac{\d\hat T^{\dagger}}{\d\lambda}\right)\hat{\cP}\hat T
+\hat T^{\dagger}\frac{\d\hat{\cP}}{\d\lambda}\hat T
+\hat T^{\dagger}\hat{\cP}\left(\frac{\d\hat T}{\d\lambda}\right),
\l{diffcong}
\f
and contracting with $\_\Psi_\Ie^{\dagger}$ and $\_\Psi_\Ie$, we find that the middle
term gives $\_\Psi^{\dagger}(\d\hat{\cP}/\d\lambda)\_\Psi$ at once, because
$\_\Psi=\hat T\_\Psi_\Ie$. The two remaining terms vanish. The argument is given in full, because it is what extends
the validity of the resulting identity beyond the dispersion surface and, as will be seen,
beyond the assumption of negligible loss.

The essential observation is that the electromagnetic block of $\hat T$ is the constant
identity, Eq.~\r{Tdef}, so that
\e
\frac{\d\hat T}{\d\lambda}
=\begin{pmatrix}0\\[2pt] \d\hat T_\Ia/\d\lambda\end{pmatrix} ,
\l{dT}
\f
that is, $\d\hat T/\d\lambda$ has a vanishing electromagnetic block. On the other hand, by
construction $\hat{\cP}\hat T=(\hat{\cS},0)^{\mathsf T}$, so that
\e
\hat{\cP}\hat T\_\Psi_\Ie
=\begin{pmatrix}\hat{\cS}\_\Psi_\Ie\\[2pt] 0\end{pmatrix}
=\begin{pmatrix}\_F_\Ie\\[2pt] 0\end{pmatrix}
\l{PT}
\f
has a vanishing {\em internal} block. The first term of Eq.~\r{diffcong} therefore pairs
the row vector $\_\Psi_\Ie^{\dagger}\,\d\hat T^{\dagger}/\d\lambda$, whose electromagnetic
entries vanish by Eq.~\r{dT}, with the column vector~\r{PT}, whose internal entries vanish;
the product is zero. Notice that no property of $\hat{\cP}$ has been used here, and that
$\hat{\cS}\_\Psi_\Ie$ need not vanish: the cancellation is purely a matter of the block
structure, and it holds for a driven field off the dispersion surface.

The third term requires one further step. We have
\e
\_\Psi_\Ie^{\dagger}\hat T^{\dagger}\hat{\cP}
=\left(\hat{\cP}^{\dagger}\hat T\_\Psi_\Ie\right)^{\dagger} ,
\f
where, for real $\_k$, $\hat{\cP}=\hat{\cH}-\ii\mathsf G$ with $\hat{\cH}$ Hermitian, so
that $\hat{\cP}^{\dagger}=\hat{\cP}+2\ii\mathsf G$. Since $\hat T\_\Psi_\Ie=\_\Psi$,
Eq.~\r{PT} gives
\e
\hat{\cP}^{\dagger}\hat T\_\Psi_\Ie
=\begin{pmatrix}\_F_\Ie\\[2pt] 0\end{pmatrix}+2\ii\,\mathsf G\_\Psi ,
\f
whose internal block is $2\ii(\mathsf G\_\Psi)_\Ia
=2\ii(\mathsf G_{\Ia\Ie}\_\Psi_\Ie+\mathsf G_{\Ia\Ia}\_\Psi_\Ia)$. The third term of
Eq.~\r{diffcong} pairs the Hermitian conjugate of this vector with
$(\d\hat T/\d\lambda)\_\Psi_\Ie$, whose electromagnetic entries vanish by Eq.~\r{dT}; it is
therefore zero for every field if the dissipation matrix has no blocks involving the
internal variables, $\mathsf G_{\Ia\Ie}=\mathsf G_{\Ia\Ia}=0$ (and hence, $\mathsf G$ being
symmetric, $\mathsf G_{\Ie\Ia}=0$), and not in general otherwise.

Collecting the two results, {\em the derivative identity requires $\_k$ to be real and the
  dissipation, if any, to be confined to the electromagnetic block of $\mathsf G$}, but it
does {\em not} require the field to satisfy a dispersion relation or the medium to be {\em completely} lossless. A lossless medium,
$\mathsf G=0$, is the simplest case, but not the only one. Dissipation in the
electromagnetic block, $\mathsf G_{\Ie\Ie}\ne0$, such as the dispersionless ohmic conductivity of a host,
is admissible however strong it is: it leaves $\hat T$ unchanged, adds the constant term
$-\ii\mathsf G_{\Ie\Ie}$ to $\hat{\cS}$, and therefore drops out of every derivative
$\d\hat{\cS}/\d\lambda$. What invalidates the identity is dissipation carried by the
variables that are eliminated (which is also discussed in Sec.~\ref{sec:scope}). Therefore, under the stated applicability conditions,
\e
\_\Psi_\Ie^{\dagger}\frac{\d\hat{\cS}}{\d\lambda}\_\Psi_\Ie
=\_\Psi^{\dagger}\frac{\d\hat{\cP}}{\d\lambda}\_\Psi .
\l{key}
\f
Equation~\r{key} relates the parametric derivatives of the reduced, purely Maxwellian
problem to similar derivatives in the problem involving the full state vector. Consequently
the AG energy density and power flux obtained from it below hold for every medium of the
class considered in which no dissipation is associated with the internal variables, driven
at any real $\o$ and $\_k$, and not merely for its eigenmodes. By contrast, the
congruence~\r{congruence} itself holds for arbitrary dissipation, and this is what makes the
heating-rate identity of Sec.~\ref{sec:heat} valid without restriction on the loss.

This result clarifies the range of applicability of the AG formulas, usually regarded as
restricted to media with negligible loss, which are derived next from the key derivative
identity of Eq.~\r{key}. What matters is not whether the medium is lossy,
but where the loss resides. By Eq.~\r{passivity}, the condition
$\mathsf G_{\Ia\Ie}=\mathsf G_{\Ia\Ia}=0$ means that the anti-Hermitian part of
$\o\dyad{\cal M}$ equals the constant matrix $\mathsf G_{\Ie\Ie}$: the loss is local and
instantaneous in the fields themselves, and the anti-Hermitian part of the material
operator comes entirely from the term $\ii\mathsf G_{\Ie\Ie}/\o$ that it contains, just as a dispersionless host
conductivity $\sigma$ adds $\ii\sigma/\o$ to the permittivity. Such loss, however strong,
is compatible with Eqs.~\r{AGenergy} and~\r{AGflux}, evaluated with the complete material
operator; since the constant term drops out of the derivatives, only the Hermitian part of
$\dyad{\cal M}$ contributes to them. Dissipation in the dispersive, eliminated degrees of
freedom is in general not compatible with them. Evaluating the third term of
Eq.~\r{diffcong} without assuming $\mathsf G_{\Ia\Ie}=\mathsf G_{\Ia\Ia}=0$ gives the error
of the identity~\r{key} in closed form, still for real $\_k$,
\e
\_\Psi_\Ie^{\dagger}\frac{\d\hat{\cS}}{\d\lambda}\_\Psi_\Ie
-\_\Psi^{\dagger}\frac{\d\hat{\cP}}{\d\lambda}\_\Psi
=-2\ii\,(\mathsf G\_\Psi)_\Ia^{\dagger}\,\frac{\d\hat T_\Ia}{\d\lambda}\,\_\Psi_\Ie .
\l{AGerror}
\f
Its right-hand side vanishes when the dissipation is confined to the electromagnetic block
and measures the departure from the identity otherwise; what this implies for the AG
formulas is taken up in Sec.~\ref{sec:validity}.

\subsection{Energy density}
\label{sec:energy}

Take $\lambda=\o$ in Eq.~\r{key}. From Eq.~\r{Pdef}, $\d\hat{\cP}/\d\o=-\mathsf M$, whereas
from Eq.~\r{Mcaldef} the $\dyad K$ block is $\o$-independent, so
$\d\hat{\cS}/\d\o=-\d(\o\dyad{\cal M})/\d\o$. Substituting into Eq.~\r{key} and using
Eq.~\r{avgs},
\e
\avg{W}=\frac14\_\Psi_\Ie^{\dagger}\frac{\d(\o\dyad{\cal M})}{\d\o}\_\Psi_\Ie ,
\l{AGenergy}
\f
the AG energy density in general bianisotropic form, which is a generalization of
Brillouin's formula for an isotropic medium~\cite{Brillouin1960,LL8}; for media with a local
response, a formula of this form was established in Ref.~\cite{Welters2014} from passivity
and analyticity, and see also Ref.~\cite{Yaghjian2018} for lossless continua with spatial
dispersion. Note that the presence of spatial dispersion
does not alter the form of this relation. For a nonmagnetic spatially dispersive dielectric,
$\dyad{\cal M}={\rm diag}(\M_0\dyad I,\dyad\E)$, and Eq.~\r{AGenergy} reduces to
\e
\avg{W}=\frac{\M_0|\_H|^2}{4}
+\frac14\_E^{*}\.\frac{\d(\o\dyad\E)}{\d\o}\.\_E ,
\l{AGenergyuniaxial}
\f
the form used in the numerical examples of Sec.~\ref{sec:hydroex} and of \SMref.

\subsection{Power flux}
\label{sec:flux}

Take now $\lambda=k_j$. Since $\hat{\cA}$ is linear in $\_k$ and neither $\mathsf M$ nor
$\mathsf G$ depends on $\_k$, $\d\hat{\cP}/\d k_j=\mathsf L_j$, and the right-hand side of
Eq.~\r{key} equals $4\avg{S_j}$ by Eq.~\r{avgs}. For the left-hand side we need the
derivative of the $\dyad K$ block; since $\dyad K\.\_a=\_k\x\_a$, we have
$(\d\dyad K/\d k_j)\.\_a=\^x_j\x\_a$, and therefore
\e
\_\Psi_\Ie^{\dagger}\frac{\d\hat{\cS}}{\d k_j}\_\Psi_\Ie
=\_H^{*}\.(\^x_j\x\_E)-\_E^{*}\.(\^x_j\x\_H)
-\o\,\_\Psi_\Ie^{\dagger}\frac{\d\dyad{\cal M}}{\d k_j}\_\Psi_\Ie .
\f
The two cross-product terms combine into the ordinary Poynting vector: using the cyclic identity
$\_a\.(\_b\x\_c)=\_b\.(\_c\x\_a)$ and
$\_H\x\_E^{*}=-\left(\_E\x\_H^{*}\right)^{*}$,
\e
\_H^{*}\.(\^x_j\x\_E)-\_E^{*}\.(\^x_j\x\_H)
=2\Re\left(\_E\x\_H^{*}\right)_j .
\l{vectorid}
\f
Collecting the terms, we obtain
\e
\avg{\_S}=\tfrac12\Re(\_E\x\_H^*)
-\frac{\o}{4}\sum_j\^x_j\left[\_\Psi_\Ie^{\dagger}\frac{\d\dyad{\cal M}}{\d k_j}\_\Psi_\Ie\right],
\l{AGflux}
\f
the AG power flux~\cite{Agranovich1966}, here written for general spatially dispersive
bianisotropic media; for lossless continua with spatial dispersion it was obtained in
Ref.~\cite{Yaghjian2018}. Unlike Eq.~\r{AGheat}, Eqs.~\r{AGenergy} and \r{AGflux} have been
derived here under the condition that no dissipation be associated with the eliminated
internal variables, which holds in particular for a lossless medium. That condition,
$\mathsf G_{\Ia\Ie}=\mathsf G_{\Ia\Ia}=0$, is sufficient rather than necessary, and
Eq.~\r{AGerror} decides the general case. A second and independent sufficient condition
applies to the flux alone: if $\mathsf L_j$ has no entry involving the internal
variables, $(\mathsf L_j)_{\Ia\Ia}=(\mathsf L_j)_{\Ie\Ia}=0$, those variables transport no
power, $\avg{S_j}$ of Eq.~\r{avgs} reduces to the ordinary Poynting term and
$\d\dyad{\cal M}/\d k_j$ vanishes with it, so that Eq.~\r{AGflux} holds whatever the
internal damping, as the Lorentz, Drude and magnetized-plasma models of
Table~\ref{tab:examples} illustrate, even though Eq.~\r{AGenergy} does not.

\subsection{What do derivatives in AG formulas measure?}
\label{sec:dictionary}

\begin{figure*}[t]
\centering
\begin{tikzpicture}[
   font=\small,
   box/.style={draw=black!55,rounded corners=2.5pt,line width=0.5pt,
               inner sep=4pt,align=center,anchor=north},
   lbl/.style={font=\footnotesize}]
\node[box,text width=0.26\textwidth] (L) at (0,0) {%
  \textbf{Extended state space}\\[2pt]
  $\_\Psi=(\_\Psi_\Ie,\_\Psi_\Ia)^{\mathsf T}$\\[2pt]
  $W=\tfrac12\_\Psi^{\mathsf T}\mathsf M\_\Psi$\\[2pt]
  $S_j=\tfrac12\_\Psi^{\mathsf T}\mathsf L_j\_\Psi$\\[2pt]
  $Q=\_\Psi^{\mathsf T}\mathsf G\_\Psi$\\[2pt]
  {\footnotesize$\mathsf M,\ \mathsf L_j,\ \mathsf G$ constant:\\ no $\o$ or $\_k$ anywhere}};
\node[box,text width=0.48\textwidth] (R) at (0.575\textwidth,0) {%
  \textbf{Maxwellian description}\\[2pt]
  $\_\Psi_\Ie=(\_H,\_E)^{\mathsf T}$, operator $\dyad{\cal M}(\o,\_k)$\\[3pt]
  $\displaystyle\avg{W}=\tfrac14\_\Psi_\Ie^{\dagger}
     \frac{\d(\o\dyad{\cal M})}{\d\o}\_\Psi_\Ie$
  \qquad
  $\displaystyle\avg{Q}=\frac{\o}{2}\_\Psi_\Ie^{\dagger}
     \frac{\dyad{\cal M}-\dyad{\cal M}^{\dagger}}{2\ii}\_\Psi_\Ie$\\[3pt]
  $\displaystyle\avg{\_S}=\tfrac12\Re(\_E\x\_H^*)
     -\frac{\o}{4}\sum_j\^x_j\,\_\Psi_\Ie^{\dagger}
     \frac{\d\dyad{\cal M}}{\d k_j}\_\Psi_\Ie$};
\coordinate (mid) at ($(L.north east)!0.5!(R.north west)$);
\draw[-{Latex[length=6pt]},line width=0.9pt]
   ($(L.north east)+(2pt,-0.95cm)$) -- ($(R.north west)+(-2pt,-0.95cm)$);
\node[lbl,align=center,above=1pt] at ($(mid)+(0,-0.95cm)$)
   {eliminate $\_\Psi_\Ia$};
\node[lbl,align=center,below=3pt] at ($(mid)+(0,-0.95cm)$)
   {Schur\\complement\\[1pt]$\hat{\cS}=\hat T^{\dagger}\hat{\cP}\hat T$};
\coordinate (ctr) at ($(L.west)!0.5!(R.east)$);
\node[lbl,anchor=north] (D) at ($(ctr |- R.south)+(0,-0.25cm)$) {%
  $\begin{array}{r@{\;\;}c@{\;\;}l@{\qquad\quad}r@{\;\;}c@{\;\;}l}
   \mathsf M\ \text{(internal entries)} &\longmapsto& \d(\o\dyad{\cal M})/\d\o &
   \mathsf G &\longmapsto& (\dyad{\cal M}-\dyad{\cal M}^{\dagger})/2\ii\\[3pt]
   \mathsf L_j\ \text{(internal entries)} &\longmapsto& \d\dyad{\cal M}/\d k_j &
   \mathsf L_0 &\longmapsto& \text{shapes $\dyad{\cal M}$; no term of its own}
  \end{array}$};
\end{tikzpicture}
\caption{The two descriptions of a medium of the class considered here, and the map between
  them. In the extended state space the energy density, the power flux and the heating rate
  are quadratic forms whose matrices carry no dependence on $\o$ or $\_k$. Eliminating the
  internal variables produces the Maxwellian description, in which the cycle averages of the
  same quantities are expressed in terms of $\o$- and $\_k$-derivatives of the material
  operator. The correspondence at the bottom
  summarizes Secs.~\ref{sec:heat}--\ref{sec:flux}: the internal entries of $\mathsf M$
  and $\mathsf L_j$ reappear as the dispersive corrections to the energy and to the flux, the
  whole of $\mathsf G$ as the anti-Hermitian part of $\dyad{\cal M}$, whereas the
  antisymmetric couplings $\mathsf L_0$, which exchange power without storing or dissipating
  it, determine $\dyad{\cal M}$ but contribute no term of their own. The correspondence
  holds at the level of the quadratic forms, Eqs.~\r{AGheat} and~\r{key}, evaluated on the
  state determined by the fields, under the conditions of Table~\ref{tab:hypotheses}.}
\label{fig:schematic}
\end{figure*}
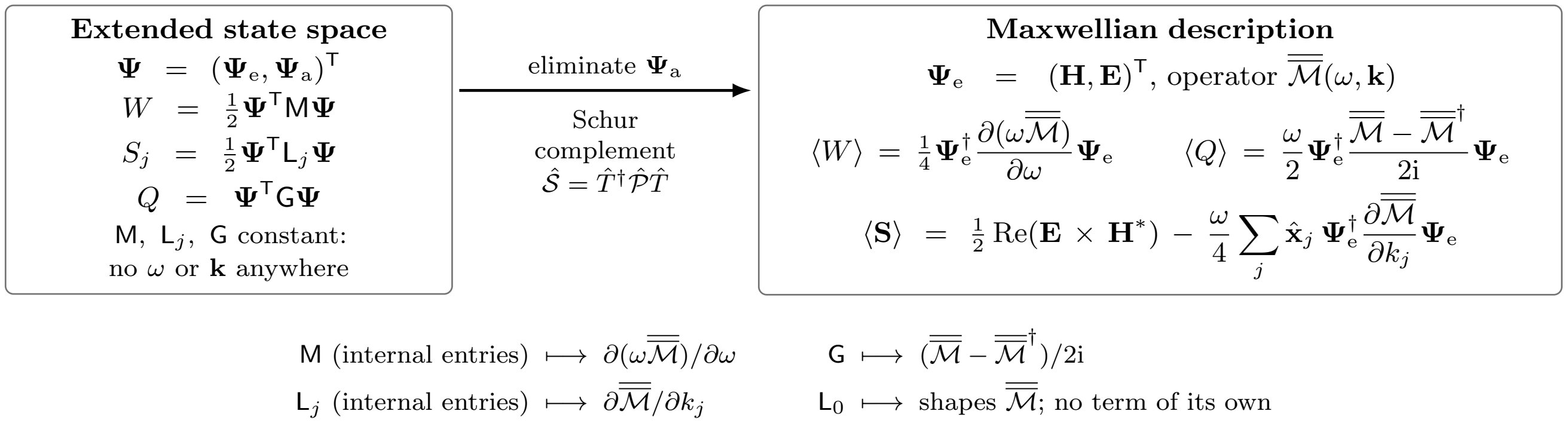

The question in the title of this subsection can now be answered explicitly. In the extended state space the energy and the flux are
the forms~\r{quadratic1}, whose matrices carry no dependence on $\o$ or $\_k$. {\em Elimination
of the internal state variables does not discard the information contained in these
matrices; it transfers it into the Maxwellian material operator.} The correspondence is collected at
the bottom of Fig.~\ref{fig:schematic}: by Eq.~\r{key}, the entries of $\mathsf M$ belonging to the
internal variables reappear in $\d(\o\dyad{\cal M})/\d\o$ and the entries of $\mathsf L_j$
that transport energy through those variables reappear in $\d\dyad{\cal M}/\d k_j$, while by
Eq.~\r{AGheat} the entries of $\mathsf G$ reappear in the anti-Hermitian part of
$\dyad{\cal M}$. The derivative corrections of the AG theory are therefore a consequence of
the elimination of the internal degrees of freedom, rather than independent constitutive
postulates or artifacts of a particular homogenization method.

The correspondence can be written as an exact identity. Suppose, as in all the examples of
Table~\ref{tab:examples} and in the wire medium of \SMref, that neither $\mathsf M$ nor $\mathsf L_j$ has blocks coupling the fields to the
internal variables, $\mathsf M_{\Ie\Ia}=0$ and $(\mathsf L_j)_{\Ie\Ia}=0$. The right-hand
side of Eq.~\r{key} then splits into an electromagnetic and an internal part. For
$\lambda=\o$ the electromagnetic part is $-\_\Psi_\Ie^{\dagger}\mathsf M_{\Ie\Ie}\_\Psi_\Ie$,
and for $\lambda=k_j$ it is the ordinary Poynting term $2\Re(\_E\x\_H^*)_j$ of
Eq.~\r{vectorid}. Removing these from both sides leaves
\begin{align}
\frac14\_\Psi_\Ie^{\dagger}\left[\frac{\d(\o\dyad{\cal M})}{\d\o}-\mathsf M_{\Ie\Ie}\right]
\_\Psi_\Ie
&=\frac14\_\Psi_\Ia^{\dagger}\mathsf M_{\Ia\Ia}\_\Psi_\Ia ,
\l{Wattrib}\\
-\frac{\o}{4}\,\_\Psi_\Ie^{\dagger}\frac{\d\dyad{\cal M}}{\d k_j}\_\Psi_\Ie
&=\frac14\_\Psi_\Ia^{\dagger}(\mathsf L_j)_{\Ia\Ia}\_\Psi_\Ia ,
\l{Sattrib}
\end{align}
where $\_\Psi_\Ia$ is the internal state set by the fields, Eq.~\r{Tdef}. In words: the
dispersive part of the AG energy density is exactly the cycle-averaged energy stored in the
eliminated variables, and the dispersive part of the AG flux is exactly the power they
transport. That the wires of a wire medium carry part of the power, and that the cycle
average of this part reproduces the AG correction, was observed for that medium in
Ref.~\cite{Silveirinha2012}; Eqs.~\r{Wattrib} and~\r{Sattrib} are the general statement, and
they identify, in any medium of this class, which internal variable carries which part. Here $\mathsf M_{\Ie\Ie}$ is the metric of the host [${\rm diag}(\M_0\dyad
I,\E_0\dyad I)$ for the standard media of Table~\ref{tab:examples}, ${\rm diag}(\M_0\dyad
I,\E_h\dyad I)$ for the wire medium], and $\d(\o\dyad{\cal M})/\d\o-\mathsf M_{\Ie\Ie}$ is therefore precisely the part of
Brillouin's correction that goes beyond the host. When $\mathsf M$ or $\mathsf L_j$ does couple
the two groups of variables, the totals~\r{AGenergy} and~\r{AGflux} remain valid, but the
cross terms prevent a unique attribution of the energy and flux to either group.

The bianisotropic $\Omega$-medium treated in Sec.~S1 of \SMref\ is an example: there the
dispersive correction to the energy density comprises the energy of the eliminated degrees
of freedom together with the energy of their interaction with the electromagnetic field.
With $\_B/\M_0$ in place of $\_H$ as the field variable, the same medium is described by a
spatially dispersive permittivity, and the interaction term moves from the energy density to
the power flux, $\_E\x\_H=\_E\x\_B/\M_0-\_E\x\_M$ (Sec.~S1\,D of \SMref).

The correspondence also says which internal variable is responsible for which correction. In
the hydrodynamic metal of Sec.~\ref{sec:examples}, a
transverse wave has $\varrho=0$, and the kinetic entry of $\mathsf M$ alone accounts for the
difference between $\d(\o\E_{\rm T})/\d\o$ and $\E_0$; in a longitudinal wave the
compressional entry adds to it, and $(\mathsf L_j)_{(\_J,\varrho)}$, which carries
the pressure-driven flux $\beta^2\varrho\_J/(\E_0\o_p^2)$, is the only source of the
wave-vector dependence of $\E_{\rm L}$. In such a wave $\_H=0$, so that the ordinary Poynting
term of Eq.~\r{AGflux} vanishes and this correction gives the entire flux
(Sec.~\ref{sec:hydroex}).

\section{Free modes of a lossless medium}
\label{sec:velocity}

\subsection{Hermitian eigenvalue problem and transport velocity}
\label{sec:eigen}

For a source-free ($\_F_{\rm ext}=0$) and lossless ($\mathsf G=0$) medium and real $\_k$,
Eq.~\r{Peq} reduces to the generalized eigenvalue problem
\e
\hat{\cA}(\_k)\_\Psi =\o\,\mathsf M\_\Psi .
\l{eigen}
\f
By conditions (i)--(ii) and Eq.~\r{symmetry}, $\hat{\cA}$ is Hermitian for real $\_k$ and
$\mathsf M$ is positive definite: the spectrum $\o(\_k)$ is automatically real, which is the
spectral expression of losslessness, and modes of distinct branches are orthogonal with
respect to the $\mathsf M$-inner product, $\_\Psi_1^{\dagger}\mathsf M\_\Psi_2=0$.

Let $\_\Psi$ be an eigenvector of Eq.~\r{eigen} with a simple eigenvalue $\o(\_k)$, so that
the branch is differentiable in $\_k$. Differentiating Eq.~\r{eigen} with respect to $k_j$
and multiplying from the left by $\_\Psi^{\dagger}$, we get
\e
\_\Psi^{\dagger}\frac{\d\hat{\cA}}{\d k_j}\_\Psi
+\_\Psi^{\dagger}\hat{\cA}\frac{\d\_\Psi}{\d k_j}
=\frac{\d\o}{\d k_j}\,\_\Psi^{\dagger}\mathsf M\_\Psi
+\o\,\_\Psi^{\dagger}\mathsf M\frac{\d\_\Psi}{\d k_j} .
\f
Because $\hat{\cA}$ is Hermitian and $\o$ is real, Eq.~\r{eigen} implies
$\_\Psi^{\dagger}\hat{\cA}=\o\_\Psi^{\dagger}\mathsf M$, so the two terms containing
$\d\_\Psi/\d k_j$ cancel and the derivative of the eigenvector drops out. Since $\hat{\cA}$
is {\em linear} in $\_k$, one has $\d\hat{\cA}/\d k_j=\mathsf L_j$ exactly, and this
Hellmann--Feynman step gives
\e
\_v_{\rm g}\equiv\frac{\d\o}{\d\_k}
=\sum_j\frac{\_\Psi^{\dagger}\mathsf L_j\_\Psi}{\_\Psi^{\dagger}\mathsf M\_\Psi}\,\^x_j
=\frac{\avg{\_S}}{\avg{W}} ,
\l{vg}
\f
where Eq.~\r{avgs} has been used in the last step and no normalization of $\_\Psi$ is
required, the ratio being homogeneous of degree zero in the field amplitude.

The energy transport velocity therefore equals the group velocity for every lossless medium
in the class considered here. The equality itself is not new. For bianisotropic media with a
local response it follows from passivity and the analytic properties of the response
functions, with the ordinary Poynting flux~\cite{Welters2014}; that argument, as its authors
note, does not extend to a general nonlocal response. For a lossless continuum with spatial
dispersion it was established with the flux that contains the wave-vector derivative of the
permittivity~\cite{Yaghjian2018}. Here it follows in one line from the structure of the
pencil, without recourse to the analytic properties of the response---indeed without the
material operator at all, since both $\avg{\_S}$ and $\avg{W}$ are evaluated in the extended
state space. It also serves as
a nontrivial consistency check on $W$ and $\_S$ of
Eq.~\r{quadratic1}: an arbitrary redistribution of the terms of the conservation law between
a stored and a transported part would not, in general, pass it.

This result has the narrowest scope of all the results of this
paper: it concerns an eigenmode of a source-free lossless medium. Its {\em derivation},
however, is the least demanding of all, and in particular it is independent of everything in
Secs.~\ref{sec:elimination}--\ref{sec:imprints}: it involves only the pencil, and it uses no
elimination, no Schur complement and no property of the material operator. The equality of
the energy transport and group velocities is therefore not a consequence of the dispersive
corrections, and it holds even at the constitutive poles, where the effective material
description does not exist.

\subsection{Integral balances for a free field}
\label{sec:integral}

The identities of Sec.~\ref{sec:imprints} hold for a single plane wave. For a free field in
the same lossless medium they carry over, with no quasi-monochromatic approximation, to two
integral characteristics, and in both cases by orthogonality. Let $\_F_{\rm ext}=0$ and
$\mathsf G=0$, and let the field be expanded in the modes of Eq.~\r{eigen},
\e
\_\Psi(\_r,t)=\Re\sum_n\int\frac{{\rm d}^3k}{(2\pi)^3}\,a_n(\_k)\,\_\Psi_n(\_k)\,
e^{\ii(\_k\.\_r-\o_n(\_k)t)} ,
\l{modalexp}
\f
the sum running over the branches with $\o_n>0$, so that each $a_n(\_k)$ is a complex
amplitude in the sense of Sec.~\ref{sec:balance}. On integrating $W$ of Eq.~\r{quadratic1}
over all space, Parseval's theorem leaves an integral over $\_k$ in which different wave
vectors do not mix. At a given $\_k$ the field is a
combination of eigenvectors of Eq.~\r{eigen}---those of the branches above, together with the
conjugated modes of $-\_k$, which are the branches with $\o_n<0$---and every term that
couples two of them carries $\_\Psi_m^{\dagger}\mathsf M\_\Psi_n$ and vanishes. What survives
is
\e
\int W\,{\rm d}^3r=\frac14\sum_n\int\frac{{\rm d}^3k}{(2\pi)^3}
\left|a_n(\_k)\right|^2\,\_\Psi_n^{\dagger}\mathsf M\_\Psi_n ,
\l{totalW}
\f
in which each term is the cycle-averaged energy density $\avg{W}$ of Eq.~\r{avgs} of the
corresponding mode, and is therefore given by Eq.~\r{AGenergy} at $(\o_n(\_k),\_k)$; for a
single plane wave Eq.~\r{totalW} returns Eq.~\r{avgs} itself.

The flux requires a different integral, since Parseval's theorem does not apply to a
surface integral. The quantity to which it does apply is the energy transported across a
plane, $U=\int{\rm d}t\int{\rm d}x\,{\rm d}y\,S_z|_{z=0}$. Transforming in $x$, $y$ and
$t$ instead, the modes that contribute at each $(\o,k_x,k_y)$ are those of Eq.~\r{eigen}
that share that frequency and those two components of the wave vector but differ in $k_z$. Such modes are
$\mathsf L_z$-orthogonal: if $\_\Psi_1$ and $\_\Psi_2$ satisfy Eq.~\r{eigen} at the same
$\o$ and at real wave vectors $\_k_1$ and $\_k_2$ differing only in $k_z$, then
$0=\_\Psi_2^{\dagger}[\hat{\cA}(\_k_1)-\hat{\cA}(\_k_2)]\_\Psi_1
=(k_{z1}-k_{z2})\,\_\Psi_2^{\dagger}\mathsf L_z\_\Psi_1$, since $\hat{\cA}$ is Hermitian at
real $\_k$ and linear in it. The cross terms in $U$ cancel in the same way as before, and
$U$ becomes a sum of the cycle-averaged fluxes $\avg{S_z}$ of Eq.~\r{avgs}, each of them
given by Eq.~\r{AGflux}. The matrix $\mathsf L_z$ is indefinite, its sign separating the
modes that carry energy towards $+z$ from those that carry it towards $-z$. The weights in
this sum are given in Appendix~\ref{app:modal}.

Both restrictions carried by Eq.~\r{eigen} are essential. In a driven field the components at
a given $\_k$ are not eigenvectors and the cross terms between different frequencies survive,
calling for an extension of the derivative identity to two spectral points, which we leave
for future work; and for $\mathsf G\ne0$ the matrix $\hat{\cA}$ is not Hermitian and the
modes are not orthogonal. The material operator thus fixes the two integral characteristics
above exactly, but not the local $W(\_r,t)$ and $\_S(\_r,t)$, which retain the cross terms
between different $\_k$ (Sec.~\ref{sec:limitations}).

\section{Example: nonlocal metal in the hydrodynamic model}
\label{sec:hydroex}

\begin{figure*}[t]
\includegraphics[width=\textwidth]{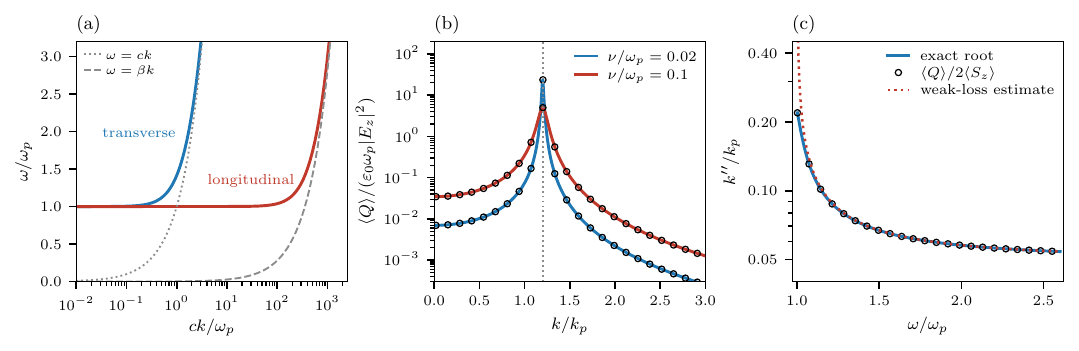}
\caption{Numerical results for the hydrodynamic model of a nonlocal metal: (a) the two
  branches of the lossless medium, as labeled, together with the light line $\o=ck$ and the
  constitutive pole $\o=\beta k$ of $\E_{\rm L}$, on a logarithmic axis because the two
  branches occupy ranges of wave number that differ by more than two decades; (b) the heating
  rate of a {\em driven} longitudinal wave versus $k$ at a fixed real $\o$, for two collision
  rates, the dotted line marking the constitutive pole; (c) the attenuation constant $k''$ of
  a freely decaying longitudinal wave versus $\o$, with the weak-loss estimate of
  Sec.~\ref{sec:attenuation} dotted. In (b) the line is the state-space heating rate and the
  circles the value computed from $\dyad\E(\o,\_k)$ alone; in (c) the line is the exact
  complex root of $\E_{\rm L}=0$ and the circles the value $\avg{Q}/(2\avg{S_z})$ furnished
  by the state-space balance~\r{attenuation}. In (b) and (c), and throughout
  Fig.~\ref{fig:hydroB}, the wave number is given in units of $k_p=\o_p/\beta$, so that none
  of the comparisons shown depends on the value of $\beta$; the parameters are listed in
  Appendix~\ref{app:hydronum}.}
\label{fig:hydroA}
\end{figure*}

We now illustrate the results obtained above on the hydrodynamic model of
Sec.~\ref{sec:examples}, which is the simplest medium of the class considered here in which
an internal variable transports energy. The results are taken in the order in which they have
been established, so that the scope narrows from one subsection to the next. Dissipation
comes first, because the heating rate of Sec.~\ref{sec:heat} is the identity that survives
it. The collision term $-\nu\_J$ of Eq.~\r{hydroJ}, which damps the electron fluid in the
same way as in the Drude case of Sec.~\ref{sec:examples}, is the only source of dissipation
in this model. After the rescaling of
Appendix~\ref{app:hydro} it becomes the only nonvanishing element of the dissipation
matrix, $\mathsf G_{(\_J,\_J)}=\nu\dyad I/(\E_0\o_p^2)$,
which is symmetric and positive semidefinite, as condition (i)
requires. The corresponding dissipation rate,
$Q=\_\Psi^{\mathsf T}\mathsf G\_\Psi=\nu\_J\.\_J/(\E_0\o_p^2)=\nu mn\,\_v\.\_v$, where $\_v$
is the fluid velocity of Appendix~\ref{app:hydro}, is the collisional loss of the electron
fluid. Notice that this damping acts on an internal variable. The present example therefore
falls on both sides of Table~\ref{tab:hypotheses}: the heating rate and the attenuation
balance survive it, whereas the AG energy density and power flux are guaranteed only for
$\nu=0$.

The collision rate enters the permittivities~\r{hydroepsLT} through $\o^2+\ii\nu\o$ in both
denominators, so that
$\Im\E_{\rm L}=\E_0\o_p^2\nu\o/|\o^2+\ii\nu\o-\beta^2k^2|^2>0$, as Eq.~\r{passivity}
requires. The lossless medium supports a transverse branch $\o^2=\o_p^2+c^2k^2$, where
$c=1/\sqrt{\E_0\M_0}$, and a longitudinal branch $\o^2=\o_p^2+\beta^2k^2$, both of which are
shown in Fig.~\ref{fig:hydroA}(a). The constitutive pole of $\E_{\rm L}$, at $\o=\beta k$,
lies on neither branch: it is the point at which the elimination of $\_J$ and $\varrho$
fails, whereas the state-space description remains regular there. Since $\beta\ll c$, the two
branches occupy ranges of wave number that differ by orders of magnitude. In what follows the
wave numbers are measured in units of $k_p=\o_p/\beta$, so that none of the longitudinal
comparisons depends on the value of $\beta$, and we take $\_k=k\^z$, so that a longitudinal
field is $\_E=E_z\^z$.

\subsection{Dissipation: heating rate and attenuation}
\label{sec:hydrolossy}

Consider first a longitudinal field driven at real $\o$ and $\_k$, which need not satisfy any
dispersion relation. In this case the elimination gives
\e
\_J=\frac{\ii\o\E_0\o_p^2}{\o^2+\ii\nu\o-\beta^2k^2}\,\_E ,
\qquad
\varrho=\frac{\_k\.\_J}{\o} ,
\l{hydrodriven}
\f
so that the state-space heating rate $\avg{Q}=\nu|\_J|^2/(2\E_0\o_p^2)$ and the
material-operator expression $\o|E_z|^2\Im\E_{\rm L}/2$ are equal identically, which
is Eq.~\r{AGheat} written out for this medium. Both are shown in Fig.~\ref{fig:hydroA}(b) for
a field driven at $\o=1.2\,\o_p$, with $k$ swept independently across the constitutive pole.
The heating rate rises sharply at the pole, because there the same $E_z$ drives a much larger
current; the peak is limited by $\nu$ alone. At that point $\E_{\rm L}$ of the lossless
medium diverges and that of the damped medium grows as $\ii\E_0\o_p^2/(\nu\o)$, whereas the
state-space description remains regular.

Figure~\ref{fig:hydroA}(c) corresponds to a different situation, that of a freely decaying
wave, in which $\o$ is real and $k$ is complex on the dispersion surface $\E_{\rm L}=0$, that
is,
\e
\beta^2k^2=\o^2-\o_p^2+\ii\nu\o .
\l{hydrocomplexk}
\f
On this surface $\_J=\ii\o\E_0\_E$ and $\varrho=\ii k\E_0E_z$, so that
\e
\avg{Q}=\frac{\nu\o^2\E_0|E_z|^2}{2\o_p^2},
\qquad
\avg{S_z}=\frac{\E_0\beta^2\o k'|E_z|^2}{2\o_p^2},
\l{hydroQS}
\f
where $k=k'+\ii k''$, and the attenuation balance $\avg{Q}=2k''\avg{S_z}$ of
Eq.~\r{attenuation} reduces to $2\beta^2k'k''=\nu\o$, which is the imaginary part of
Eq.~\r{hydrocomplexk} itself. It holds, therefore, for an arbitrary damping. Also shown in
the figure is the perturbative estimate discussed at the end of Sec.~\ref{sec:attenuation},
in which $\avg{Q}$ and $\avg{S_z}$ are evaluated with the fields of the lossless mode. It
gives $k''\simeq\nu\o/(2\beta^2k_0)$ with $\beta k_0=\sqrt{\o^2-\o_p^2}$ and fails near the
plasma edge, where $k_0\to0$ while the exact $k''$ remains finite.

\subsection{Energy density and power flux}
\label{sec:hydrofields}

\begin{figure*}[t]
\includegraphics[width=\textwidth]{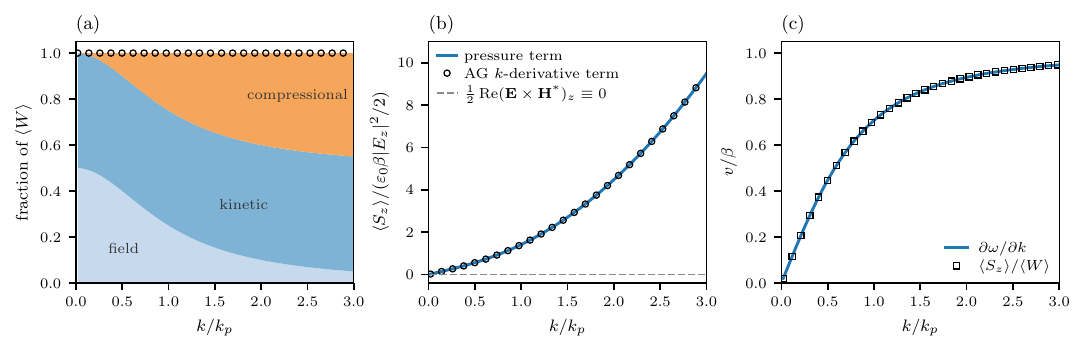}
\caption{The lossless longitudinal branch of the same model. (a) Shares of $\avg{W}$,
  Eq.~\r{hydroW}: the electric field energy and the kinetic and compressional energies of the
  electron fluid, stacked as shaded areas normalized to their sum; the circles show the AG
  energy~\r{AGenergy}, evaluated from $\dyad\E(\o,\_k)$ alone and normalized to the same sum,
  so that agreement means circles at unity. (b) The flux $\avg{S_z}$: the pressure term
  $\beta^2\Re(\varrho J_z^{*})/(2\E_0\o_p^2)$, which transports power through the electron
  fluid (line), the wave-vector-derivative term of Eq.~\r{AGflux} (circles), and the ordinary
  Poynting term, which vanishes identically because $\_H=0$. (c) Energy transport velocity:
  $\d\o/\d k$ obtained from the branch (line) and $\avg{S_z}/\avg{W}$ evaluated from the
  state-space forms (squares), in units of $\beta$.}
\label{fig:hydroB}
\end{figure*}

We now set $\nu=0$, which is the further narrowing of scope that Eqs.~\r{AGenergy}
and~\r{AGflux} require, and follow the longitudinal branch, on which $\o^2-\beta^2k^2=\o_p^2$
and, therefore, $\_J=\ii\o\E_0\_E$ and $\varrho=\ii k\E_0E_z$. A longitudinal wave has
$\_H=0$, and the cycle-averaged energy density of the state-space model consequently contains
three terms,
\begin{multline}
\avg{W}=\underbrace{\frac{\E_0|E_z|^2}{4}}_{\rm field}
+\underbrace{\frac{|\_J|^2}{4\E_0\o_p^2}}_{\rm kinetic}
+\underbrace{\frac{\beta^2|\varrho|^2}{4\E_0\o_p^2}}_{\rm compressional}\\
=\frac{\E_0|E_z|^2}{4}\left(1+\frac{\o^2+\beta^2k^2}{\o_p^2}\right) ,
\l{hydroW}
\end{multline}
whose shares are plotted in Fig.~\ref{fig:hydroB}(a). Differentiating the longitudinal
permittivity and using the branch relation, we obtain
\begin{multline}
\frac{\d(\o\E_{\rm L})}{\d\o}
=\E_0\left[1+\frac{\o_p^2(\o^2+\beta^2k^2)}{(\o^2-\beta^2k^2)^2}\right]\\
=\E_0\left(1+\frac{\o^2+\beta^2k^2}{\o_p^2}\right) .
\l{hydrodwe}
\end{multline}
The AG energy density~\r{AGenergy}, which for $\_H=0$ reduces to
$\tfrac14|E_z|^2\d(\o\E_{\rm L})/\d\o$, therefore reproduces the total energy term by term:
the unity in Eq.~\r{hydrodwe} corresponds to the field energy, and the dispersive remainder
to the kinetic and compressional energies of the electron fluid. This is Eq.~\r{Wattrib}
written for the present medium. The derivative $\d[\o(\E_{\rm L}-\E_0)]/\d\o$ thus measures the
energy of the electron fluid and nothing else.

Evaluation of the power flux leads to a stronger conclusion. Since $\_H=0$, the ordinary Poynting term
of Eq.~\r{AGflux} vanishes identically, and the whole flux is carried by
$(\mathsf L_j)_{(\_J,\varrho)}$,
\e
\avg{S_z}=\frac{\beta^2}{2\E_0\o_p^2}\Re(\varrho J_z^{*})
=\frac{\E_0\beta^2\o k}{2\o_p^2}|E_z|^2 .
\l{hydroS}
\f
On the AG side, only $\E_{\rm L}$ is differentiated, because the derivative of $\^k\^k$ has
no $zz$ component at $\_k=k\^z$, so that the wave-vector correction of Eq.~\r{AGflux} reads
\e
-\frac{\o}{4}|E_z|^2\frac{\d\E_{\rm L}}{\d k}
=\frac{\o}{4}|E_z|^2\frac{2\E_0\o_p^2\beta^2k}{(\o^2-\beta^2k^2)^2}
=\frac{\E_0\beta^2\o k}{2\o_p^2}|E_z|^2 ,
\l{hydroSag}
\f
which is the same quantity, Fig.~\ref{fig:hydroB}(b). For a transverse wave the situation is
exactly the opposite: $\varrho=0$, and the correction vanishes, because $\E_{\rm T}$ does not
depend on the wave vector while the derivative of $\^k\^k$ does not contribute for
$\_E\perp\_k$; the whole flux is then the ordinary Poynting term. Notice that for
such a wave the flux identity is satisfied trivially for an arbitrary $\nu$. In a
longitudinal wave, on the contrary, the wave-vector derivative of the permittivity is not a
correction to the power flux but the power flux itself. This is the extreme case of
Eq.~\r{Sattrib}: what the derivative measures is the pressure flux carried by
$(\mathsf L_j)_{(\_J,\varrho)}$.

Collisional damping breaks these two identities, and Eq.~\r{AGerror} gives the resulting
error. Its residual is proportional to $(\mathsf G\_\Psi)_\Ia$ and is, therefore, of the
first order in the loss, provided that the field is far from the constitutive pole. Indeed,
for a wave driven at $\o=1.7\,\o_p$ and $k=0.9\,k_p$, the mismatch of Eq.~\r{AGenergy} grows
in proportion to $\nu$ and is $7.5\x10^{-4}$, $7.5\x10^{-3}$, and $7.5\x10^{-2}$ of
$\avg{W}$ at $\nu/\o_p=10^{-3}$, $10^{-2}$, and $10^{-1}$, respectively, whereas that of
Eq.~\r{AGflux} is $1.6\x10^{-3}$, $1.6\x10^{-2}$, and $1.6\x10^{-1}$ of $\avg{S_z}$. This
discrepancy, however, is almost entirely due to the imaginary part that the loss adds to the
AG expressions and which has no counterpart in $\avg{W}$ and $\avg{S_z}$. Their real parts
still reproduce these two quantities to the second order in $\nu$, to relative errors of
$6\x10^{-7}$, $6\x10^{-5}$, and $6\x10^{-3}$ for the energy. At the pole itself, on the
other hand, the error is of the order of the energy stored in the internal variables, no
matter how small $\nu$ is, for the reason explained in Sec.~\ref{sec:validity}.

\subsection{Energy transport velocity}
\label{sec:hydrovg}

The last and narrowest of the checks is Eq.~\r{vg}, which holds only for an eigenmode of the
lossless medium. Differentiating the longitudinal branch gives $\d\o/\d k=\beta^2k/\o$, and
dividing Eq.~\r{hydroS} by Eq.~\r{hydroW} gives the same result,
\e
\frac{\avg{S_z}}{\avg{W}}
=\frac{\E_0\beta^2\o k|E_z|^2/(2\o_p^2)}{\E_0\o^2|E_z|^2/(2\o_p^2)}
=\frac{\beta^2k}{\o} ,
\f
where $\avg{W}$ has been simplified with $\o^2-\beta^2k^2=\o_p^2$. The two expressions for
the group velocity are compared in Fig.~\ref{fig:hydroB}(c), where they agree to machine
precision. For the transverse branch the
same calculation gives $c^2k/\o$, with the flux entirely in $\_E\x\_H$.

\section{Discussion}
\label{sec:scope}

\begin{table*}[t]
\caption{\label{tab:hypotheses}Conditions under which each result holds. ``Loss'': whether
  the result survives $\mathsf G\ne0$. ``Complex $\_k$'': whether it applies to an
  attenuated wave. ``Sources'': which external forces are admitted; ``$\_F_\Ie$ only'' means
  that the medium is driven through the electromagnetic fields, $\_F_\Ia=0$. ``Mode only'':
  whether $\o$ and $\_k$ must satisfy the dispersion relation, that is, whether the result
  holds only for a free mode of the medium. A dash means that the question does not arise.
  ``$\mathsf G_{\Ie\Ie}$ only'' means that loss is admitted, of any strength, provided it is
  confined to the electromagnetic block $\mathsf G_{\Ie\Ie}$, with
  $\mathsf G_{\Ia\Ie}=\mathsf G_{\Ie\Ia}=\mathsf G_{\Ia\Ia}=0$
  (Sec.~\ref{sec:derivid}); this is a sufficient condition, and Eq.~\r{AGerror} gives the
  exact criterion.}
\begin{ruledtabular}
\begin{tabular}{llcccc}
Result & Follows from & Loss & Complex $\_k$ & Sources & Mode only\\
\colrule
Conservation law, Eq.~\r{Poyntingtheorem} & Eq.~\r{symmetry} & yes & --- & any & no\\
Congruence, Eq.~\r{congruence} & block algebra & yes & yes & --- & no\\
Heating rate, Eq.~\r{AGheat} & congruence & yes & no & $\_F_\Ie$ only & no\\
Attenuation balance, Eq.~\r{attenuation} & conservation law & yes & yes & none & yes\\
Derivative identity, Eq.~\r{key} & congruence & $\mathsf G_{\Ie\Ie}$ only & no & $\_F_\Ie$ only & no\\
AG energy, Eq.~\r{AGenergy} & derivative identity & $\mathsf G_{\Ie\Ie}$ only & no & $\_F_\Ie$ only & no\\
AG flux, Eq.~\r{AGflux} & derivative identity & $\mathsf G_{\Ie\Ie}$ only\footnotemark[1] & no & $\_F_\Ie$ only & no\\
$\_v_{\rm g}=\avg{\_S}/\avg{W}$, Eq.~\r{vg} & Hellmann--Feynman & no & no & none & yes\\
\end{tabular}
\end{ruledtabular}
\footnotetext[1]{Or $(\mathsf L_j)_{\Ia\Ia}=(\mathsf L_j)_{\Ie\Ia}=0$, in which case the
  flux formula holds at any loss (Sec.~\ref{sec:flux}); the Lorentz, Drude and
  magnetized-plasma models are of this kind.}
\end{table*}

The usual phenomenological approach to the energy density and power flux in a dispersive medium is to
start from the Maxwell equations with a given material operator $\dyad{\cal M}$
and then look for suitable quadratic forms of the fields that may represent these quantities under some approximations.
Unfortunately, the electromagnetic energy balance alone does not determine such forms uniquely, and this
is the source of ambiguities mentioned in the literature. In this article, we have
proceeded in a different direction. The energy density, the power flux, and the heating
rate are first obtained from the equations of wave motion formulated in an extended state
space and then followed through the elimination of the internal variables. Along the main
chain of derivations the conditions become stricter step by step. No assumption is made
about the strength of the spatial dispersion, or the microscopic structure of the medium.
Within the extended state-space picture, the conservation law of Eq.~\r{Poyntingtheorem}
holds for arbitrary fields in an inhomogeneous medium, with any dissipation and any
external sources. After developing the necessary analytic tools in
Sec.~\ref{sec:elimination}, in Secs.~\ref{sec:dissipation} and~\ref{sec:imprints} we therefore
address the following question: under what conditions, and in what form, can the
quantities from this conservation law be expressed in terms of the constitutive parameters
of the reduced problem, that is, in terms of the Maxwellian material operator? The answers
are the heating rate of Sec.~\ref{sec:heat} and the generalized AG formulas for the stored
energy and the power flux of Secs.~\ref{sec:energy} and~\ref{sec:flux}. The conditions under
which each of them holds are collected in Table~\ref{tab:hypotheses}, which also shows the
identity from which each result follows, and are discussed in more detail below.

\subsection{Range of validity of AG formulas}
\label{sec:validity}

The heating rate~\r{AGheat} holds for
dissipation of any strength, but it requires a real frequency and wave vector and sources that act on the
electromagnetic fields alone, with $\_F_\Ia=0$. The AG energy density and power flux,
Eqs.~\r{AGenergy} and~\r{AGflux}, require, in addition, that no dissipation be associated with
the internal variables. None of these three results requires the field to be a free mode of
the medium: $\o$ and $\_k$ need not satisfy the dispersion relation. This is important
because the field of a distributed source in a uniform medium is a superposition of such
driven plane waves, and the results apply to each of them. The attenuation
balance~\r{attenuation} and the relation $\_v_{\rm g}=\avg{\_S}/\avg{W}$, in contrast,
concern free waves and hold only on the dispersion surface.

For the media considered here the AG formulas are therefore exact identities rather than
approximations for quasi-monochromatic wave packets. Each of the
hypotheses under which they have been established previously is relaxed here in a
different respect: periodicity is not required~\cite{Silveirinha2009}, the response may be
nonlocal~\cite{Welters2014}, and the wave vector enters on the same footing as the
frequency~\cite{Silveirinha2021}. What is assumed instead is the existence of a local
realization satisfying conditions (i)--(iii), which is a genuine restriction on the medium
and is discussed in Sec.~\ref{sec:limitations}. The present results are also not limited to
the lossless case, for which the spatially dispersive flux was obtained
previously~\cite{Yaghjian2018}, or to negligible loss: as shown in
Sec.~\ref{sec:derivid}, they remain exact for loss of any strength, provided that it is
confined to the electromagnetic block of $\mathsf G$, that is, acts directly on the fields,
e.g.\ through a nondispersive conductivity $\sigma$.

Outside these conditions the expressions written in terms of the Maxwellian material operator can fail,
but the extended state-space model never does, and the derivation shows exactly where such failures may occur. Four situations of this kind can be distinguished.

First, when there is dissipation associated with the internal variables, the third term of
Eq.~\r{diffcong} no longer vanishes, and the error of the derivative identity is the one
given in Eq.~\r{AGerror} of Sec.~\ref{sec:derivid}.
The error is proportional to $(\mathsf G\_\Psi)_\Ia$, that is, to the dissipation in the
internal variables, such as the collisional damping of the carriers in a Drude or
hydrodynamic metal (Sec.~\ref{sec:hydroex}) or the
resistance of the wires in the wire-medium example of \SMref. It is small when this
dissipation is weak and
the field is far from a constitutive pole. Near a pole, however, the error is comparable to
the energy stored in the internal variables, no matter how weak the loss is. The reason is
that the internal response is then limited only by the loss. One power of
$\hat{\cP}_{\Ia\Ia}^{-1}$ cancels the explicit factor $\mathsf G$ in Eq.~\r{AGerror}, so
that $(\mathsf G\_\Psi)_\Ia$ stays finite; the two further powers carried by
$\d\hat T_\Ia/\d\lambda$ make the error diverge at the same rate as the energy stored in the
internal variables, both as the inverse square of the loss, so that their ratio remains of
order unity however weak the loss is. This agrees with the known
failure of Brillouin's formula inside absorption lines~\cite{LL8}. The heating-rate
identity~\r{AGheat} is not affected by such loss and holds across the pole, as is seen in
Fig.~\ref{fig:hydroA}(b) and in Fig.~S2(b) of \SMref.

Second, near a constitutive pole the reduced description itself breaks down: in a lossless
medium the Maxwellian material operator does not exist there, and with loss it varies rapidly
near the pole (Sec.~\ref{sec:schur}).

Third, sources acting directly on the internal variables, $\_F_\Ia\ne0$, put part of the
internal state beyond the reach of any expression built from the fields and
$\dyad{\cal M}$, for the reason given in Sec.~\ref{sec:system}.

Fourth, the free waves of a lossy medium have
complex $\_k$. For them the AG formulas have not been established, and the relation that
holds is the attenuation balance~\r{attenuation}, which follows directly from the extended state-space
model.

\subsection{Model limitations and extensions}
\label{sec:limitations}

Under the conditions in which the derivative identity of Eq.~\r{key} holds, the material
operator determines both the energy density and the power flux of a single plane wave:
Eqs.~\r{AGenergy} and~\r{AGflux} express $\avg{W}$ and $\avg{\_S}$ in terms of
$\dyad{\cal M}(\o,\_k)$ alone. Any two state-space models that
satisfy these conditions and lead to the same material operator therefore give the same
energy density and power flux for each plane wave. Conversely, an expression for these
quantities that disagrees with Eqs.~\r{AGenergy} and~\r{AGflux} for a plane wave is
inconsistent with every such model.

A nonuniform field, such as a pulse, a beam, or a standing wave, is a superposition of
plane waves. Besides the contributions of the individual plane waves, its energy density
and power flux contain cross terms, which describe the beating between pairs of plane
waves and determine how the energy and the power flux are distributed in space and
time. The results obtained here separate the following two
questions that are often discussed together in the literature in connection with the
non-uniqueness of the energy density and the Poynting vector in spatially dispersive
media~\cite{Markel2008,Markel2010,Richter2010,SilveirinhaReply2010}.
The first concerns a source-driven monochromatic plane wave, and is settled by
Eqs.~\r{AGenergy} and~\r{AGflux} under the conditions stated above. For a free field in a
lossless medium the answer carries over, by Sec.~\ref{sec:integral}, to the {\em integral}
characteristics of the field: the energy of a pulse of finite extent and the energy it
transports across a plane. The second concerns the space-time distributions of the energy-
and flux-related quantities {\em in an arbitrary nonuniform time-varying field}: these are
not fixed by the Maxwellian energy balance alone, and they have to be obtained from the
spatio-temporally local state-space model introduced in this work.

When there is loss in the internal variables, the material operator may not determine the
energy even for a single plane wave. For a dissipative
system the storage function, which plays the role of the stored energy, is in general not
unique even for a minimal realization~\cite{Willems1972a}. Several energy metrics may then
be compatible with the same material operator and with conditions (i)--(ii), and the
physical one has to be selected by a model of the internal processes. The same conclusion
was reached for wire media in Ref.~\cite{Silveirinha2012}.

Furthermore, a homogenized model does not automatically have the structure of Eq.~\r{model}. As seen in
Sec.~S2\,B of \SMref\ and Appendix~\ref{app:hydro}, the equations of motion of the
internal variables must be rescaled in a particular way for $\hat{\cL}$ to be skew-adjoint,
and this rescaling fixes the energy metric. A homogenized model with this structure has a
local and instantaneous energy balance, and all the results of this paper apply to it under
the conditions of Table~\ref{tab:hypotheses}. Floquet--Bloch periodicity at the microscopic level is not required, so quasi-periodic
and amorphous media are included, provided that a local state-space model satisfying conditions
(i)--(iii) exists for them. For wire media, such a model is given by the quasistatic theory of
Ref.~\cite{Maslovski2009} in the thin-wire regime. In such media the AG formulas follow
exactly from the local model, and they are therefore as accurate, with respect to the
microscopic fields, as the model itself.

The present approach thus complements, rather than replaces, the first-principles derivation
of Ref.~\cite{Silveirinha2009}, which starts from the microscopic Maxwell equations of a
periodic structure. Our approach builds up from a state-space model with
a finite number of internal variables; for such a model $\dyad{\cal M}$ is a rational
function of $\o$ and $\_k$. Responses with branch cuts, such as the Lindhard response of an electron gas
with Landau damping, would require a continuum of internal variables, as in the Hamiltonian
and Lagrangian extensions of Refs.~\cite{Tip1998,Figotin2005,Figotin2007,Philbin2010}. The hydrodynamic model of
Sec.~\ref{sec:examples} is a truncation of this response. With a continuum of internal
variables the block algebra is formally the same, but the invertibility of
$\hat{\cP}_{\Ia\Ia}$ and the notion of minimality require a more careful treatment.

The local extended state-space formulation is also useful in two problems for which the Maxwellian material operator alone
is not sufficient. The first concerns boundaries. Because the conservation law holds at every
point of each medium, the normal component of $\_S$ must be continuous across an interface
that neither stores nor dissipates energy. This gives a necessary condition for the
additional boundary conditions of spatially dispersive media~\cite{Maslovski2010}. The second
concerns time-domain modeling. Since Eq.~\r{model} is a first-order system with an exact
instantaneous energy $W=\frac12\_\Psi^{\mathsf T}\mathsf M\_\Psi$, it can be used directly,
in the spirit of the auxiliary-differential-equation methods, to model pulses in such media
with correct energy accounting and without the quasi-monochromatic approximation.

The frequency-domain results of Secs.~\ref{sec:elimination}--\ref{sec:velocity} were derived
for plane waves in a uniform medium. In a slowly varying inhomogeneous medium they can be used
only within a local-uniformity approximation, which we have not studied. The conservation
law~\r{Poyntingtheorem}, in contrast, holds for inhomogeneous media without approximation.

The identities~\r{congruence}, \r{AGheat}, \r{key}, \r{AGenergy} and~\r{AGflux} do not
use the positive definiteness of $\mathsf M$
or the positive semidefiniteness of $\mathsf G$. These properties are needed only to
interpret $W$ as a stored energy and $Q$ as a dissipation rate, to prove that the energy is
unique and the material operator passive, and to make the spectrum of Eq.~\r{eigen} real.
The identities therefore also apply to active and non-Foster media, in which the stored energy
may become negative~\cite{Serra2023}. In such media, however, the eigenvalue
problem~\r{eigen} is self-adjoint only with respect to the indefinite $\mathsf M$-inner product, and complex eigenfrequencies, that is, instabilities, may appear. This extension
remains open.

Time-modulated media are outside the frequency-domain analysis, because their material
operator is not a function of a single frequency. The local state-space model, however, still applies. A
time dependence of $\mathsf L_0$, $\mathsf L_j$, and $\mathsf G$ that preserves Eq.~\r{symmetry} at
every instant leaves the conservation law unchanged, whereas a time-dependent $\mathsf M$ adds
to it one term, the power exchanged with the agent that drives the modulation.

Finally, only the electromagnetic block, in which Maxwell's equations fix the form of
$\dyad K$ in Eqs.~\r{Mcaldef} and~\r{Sofcal}, is specific to electromagnetism. We therefore
expect the analysis to apply also to acoustic and elastic metamaterials. There, eliminating
the internal variables produces an effective density and stiffness that depend on frequency
and wave vector, together with Willis-type coupling, the counterpart of bianisotropy.

\section{Conclusions}
\label{sec:conclusions}

We have considered linear time-invariant media whose macroscopic dynamics closes as a local,
first-order system for an extended state vector, in which the electromagnetic field is
complemented by internal variables such as polarizations, currents, or the charges of lumped
loads. In this system the energy metric is symmetric and positive definite, the dissipation
matrix is positive semidefinite, and the spatial operator is skew-adjoint. The energy density, the power
flux, and the heating rate are then quadratic forms of the extended state with matrices
independent of frequency and wave vector. The temporally and spatially
dispersive material operator appears only when the internal variables are eliminated, as the
Schur complement of the internal block of the state-space pencil.

The congruence that relates this Schur complement to the pencil is what carries the
quadratic forms over to the Maxwellian description. Through it, a fixed metric that contains
no derivatives at all reappears as the frequency and wave-vector derivatives of the material
operator, which are therefore imprints of the eliminated degrees of freedom rather than
independent constitutive postulates. The equality of the
energy transport and group velocities, known before for local passive
media~\cite{Welters2014} and for lossless spatially dispersive
continua~\cite{Yaghjian2018}, is by contrast independent of all this: it holds already in the
extended state space, before any elimination, and therefore also where the material operator
does not exist.

The main results can be summarized as follows. First, the heating rate of a field driven at
real $\o$ and $\_k$ by sources acting on the electromagnetic fields alone is obtained from
the congruence identity alone, for arbitrary dissipation, and the passivity of the
material operator is inherited from the state-space model rather than assumed. Second, the
AG energy density and power flux follow from a derivative identity whose proof uses only
the block structure of the elimination. They are exact identities rather than approximations
for quasi-monochromatic wave packets: they hold for fields driven at arbitrary real $\o$
and $\_k$, and not only for eigenmodes, and they remain exact for loss of arbitrary strength,
provided that it is carried by the electromagnetic subsystem alone and that the sources act
on the electromagnetic fields alone. Loss associated with the eliminated variables
introduces an error, Eq.~\r{AGerror}, which is small for weak loss away from constitutive
poles. Third, the dispersive parts of the AG formulas
are shown to equal, exactly, the cycle-averaged energy density and power flux of the
eliminated variables [Eqs.~\r{Wattrib} and \r{Sattrib} in Sec.~\ref{sec:dictionary},
whenever $\mathsf M$ and $\mathsf L_j$ have no blocks coupling the fields to the internal
variables, as in all the examples except the $\Omega$-medium of \SMref], which is the physical meaning
of the frequency and wave-vector derivatives. Fourth, none of this uses microscopic periodicity, weak
nonlocality, or an assumed pole structure of the material operator, whose rational form
follows from the realization instead of being postulated. Every hypothesis is stated where
it enters, and all of them are collected in Table~\ref{tab:hypotheses}. Taken together,
these results mean that, within the range that table delimits, the material operator
determines the energy
and the power flux of every plane-wave component and the integral quantities of a
superposition, such as the total energy of a pulse, whereas the local distribution of the
energy and the power flux in a nonuniform field, which the energy balance alone leaves open,
is given unambiguously by the local model.

The two worked examples show that the dispersive terms need not be small corrections. In the
hydrodynamic metal of Sec.~\ref{sec:hydroex} a longitudinal wave carries no magnetic field,
so that the ordinary Poynting term vanishes and the wave-vector derivative of the
permittivity gives the whole power flux, whereas the frequency derivative accounts for the
kinetic and compressional energies of the electron fluid, which dominate $\avg{W}$ away from
the plasma edge. The capacitively loaded wire medium worked out in \SMref\ leads to the same
conclusion for a metamaterial with lumped loads. On the branch that reduces to a transverse
electromagnetic wave in the unloaded limit, a large and growing part of the stored energy and
of the transported power resides in the wire currents, the additional potentials, and the load
charges, none of which appears in the effective permittivity; yet the AG formulas evaluated
from the effective permittivity alone reproduce both totals exactly within the model.

Beyond the plane-wave energy formulas, the local formulation provides a structural criterion
for homogenized models of metamaterials, periodic or not, with a well-defined energy balance;
it imposes an energy constraint on additional boundary conditions and supplies an exact energy
functional for time-domain modeling. It also remains applicable where the material operator
is singular or is not a function of a single frequency. Time-modulated and active media, and
responses that require a continuum of internal variables, are natural directions for future
work.

\begin{acknowledgments}
  The authors acknowledge the use of Claude Opus~5 (Anthropic) as an assistive tool for
  drafting symbolic derivations and numerical codes and for language editing. All
  derivations and results were verified by the authors, who take full responsibility for
  the content. S.M. acknowledges financial support through FCT -- Funda\c{c}\~{a}o para a
  Ci\h{e}ncia e a Tecnologia, I.P., by national funds, and, when eligible, co-funded by EU funds under
  project UID/50008/2025 -- Instituto de Telecomunica\c{c}\~oes, with DOI identifier --
  {\tt https://doi.org/10.54499/UID/50008/2025}.
\end{acknowledgments}

\appendix

\section{Skew-adjointness and cycle averages}
\label{app:adjoint}

For real square-integrable vector fields $\_a$ and $\_b$ that decay at infinity, the natural
inner product is $\langle\_a,\_b\rangle=\int\_a^{\mathsf T}\_b\,{\rm d}V$. Integrating by
parts,
\e
\int\_a^{\mathsf T}\mathsf L_j\frac{\d\_b}{\d x_j}\,{\rm d}V
=-\int\left(\mathsf L_j^{\mathsf T}\frac{\d\_a}{\d x_j}\right)^{\mathsf T}\!\!\_b\,{\rm d}V ,
\l{byparts}
\f
where the boundary term has been dropped. Hence
\e
\hat{\cL}^{\dagger}=-\sum_j\mathsf L_j^{\mathsf T}\frac{\d}{\d x_j}+\mathsf L_0^{\mathsf T} ,
\l{adjoint}
\f
and the requirement $\hat{\cL}^{\dagger}=-\hat{\cL}$, compared term by term, is equivalent
to Eq.~\r{symmetry}: the derivative terms require $\mathsf L_j^{\mathsf T}=\mathsf L_j$ and the
zeroth-order term requires $\mathsf L_0^{\mathsf T}=-\mathsf L_0$.
Notice that the derivative operator $\d/\d x_j$ is itself skew-adjoint, which is why
the {\em symmetric} matrices $\mathsf L_j$ and the {\em antisymmetric} matrix $\mathsf L_0$ enter
on the same footing.

For the cycle averages, write
$\_\Psi(\_r,t)=\Re[\tilde{\_\Psi}(\_r)e^{-\ii\o t}]$. For any real symmetric matrix
$\mathsf B$,
\e
\_\Psi^{\mathsf T}\mathsf B\_\Psi
=\tfrac14\left(
\tilde{\_\Psi}^{\mathsf T}\mathsf B\tilde{\_\Psi}\,e^{-2\ii\o t}
+\tilde{\_\Psi}^{\dagger}\mathsf B\tilde{\_\Psi}^{*}\,e^{2\ii\o t}
+2\,\tilde{\_\Psi}^{\dagger}\mathsf B\tilde{\_\Psi}
\right),
\f
and the oscillating terms average to zero over a period, leaving
$\avg{\_\Psi^{\mathsf T}\mathsf B\_\Psi}
=\frac12\tilde{\_\Psi}^{\dagger}\mathsf B\tilde{\_\Psi}$, which applied to
Eqs.~\r{quadratic1} and \r{quadratic2} gives Eq.~\r{avgs}. The tilde is omitted throughout
the main text. For a plane wave with complex $\_k$ the amplitude $\tilde{\_\Psi}$ carries the
factor $e^{-\_k''\.\_r}$, and the same result gives position-dependent averages proportional
to $e^{-2\_k''\.\_r}$, as used in Sec.~\ref{sec:attenuation}.

\section{Reduction of the hydrodynamic model to first-order form}
\label{app:hydro}

In the linearized hydrodynamic model of a metal, the electron fluid of equilibrium density
$n$ obeys $m\,\d\_v/\d t=-e\_E-n^{-1}\D p-m\nu\_v$ with the pressure perturbation $p=m\beta^2\delta n$, where
$\delta n$ is the density perturbation, $\beta$ the hydrodynamic velocity and $\nu$ the
collision rate. With
$\_J=-en\_v$ and the induced charge density $\varrho=-e\delta n$, this is
$\d\_J/\d t=\E_0\o_p^2\_E-\beta^2\D\varrho-\nu\_J$, and the continuity equation is
$\d\varrho/\d t=-\D\.\_J$, which are Eqs.~\r{hydro}, with $\_J$ and $\varrho$ related to
the polarization by Eq.~\r{hydroP}.

Writing the two equations as block rows of Eq.~\r{model} and rescaling the $\_J$ row by
$a_{\_J}$ and the $\varrho$ row by $a_\varrho$, the pair of entries coupling $\_E$ and $\_J$
is antisymmetric only for $a_{\_J}=1/(\E_0\o_p^2)$, whereas the pair coupling $\_J$ and
$\varrho$ belongs to $\mathsf L_j$ and, $\d_j$ being skew-adjoint, must be {\em symmetric},
which requires $a_\varrho=a_{\_J}\beta^2$. This fixes the entries quoted in
Table~\ref{tab:examples} and gives
\e
W=\frac{\M_0|\_H|^2}{2}+\frac{\E_0|\_E|^2}{2}
+\frac{|\_J|^2}{2\E_0\o_p^2}+\frac{\beta^2\varrho^2}{2\E_0\o_p^2} ,
\f
whose last two terms are $\frac12 mn|\_v|^2$ and $\frac12 m\beta^2(\delta n)^2/n$, the kinetic and
compressional energies of the electron fluid.

The elimination of auxiliary variables can be done as follows. For a plane wave, Eqs.~\r{hydro} become
$-\ii\o\_J=\E_0\o_p^2\_E-\ii\beta^2\_k\varrho-\nu\_J$ and $-\ii\o\varrho=-\ii\_k\.\_J$. The second
gives $\varrho=\_k\.\_J/\o$, and substitution into the first leaves
\e
-\frac{\ii}{\o}\left[(\o^2+\ii\nu\o)\_J-\beta^2\_k(\_k\.\_J)\right]=\E_0\o_p^2\_E .
\f
The transverse part of the current, $\_k\.\_J_{\rm T}=0$, gives
$\_J_{\rm T}=\ii\o\E_0\o_p^2\_E_{\rm T}/(\o^2+\ii\nu\o)$, and the longitudinal part gives
$\_J_{\rm L}=\ii\o\E_0\o_p^2\_E_{\rm L}/(\o^2+\ii\nu\o-\beta^2k^2)$. With $\_P=\_J/(-\ii\o)$ and
$\_D=\E_0\_E+\_P$ this is Eq.~\r{hydroepsLT}. The same result follows from Eq.~\r{schur};
here we have simply performed the necessary substitutions explicitly, without resorting to
the block algebra.

\section{Modal expansion and the integral balances}
\label{app:modal}

The sum in the expansion~\r{modalexp} runs over the dynamical branches of Eq.~\r{eigen},
and over those alone. Besides these branches, the same eigenvalue problem has solutions
with $\o=0$, of three kinds: static longitudinal fields, which the divergence constraints of
Maxwell's equations exclude; static states of the internal variables, such as the
pre-charged capacitors of the loaded wire medium of \SMref; and, in a lossless medium with
free carriers, static magnetic fields sustained by persistent currents. These states carry no flux and are
$\mathsf M$-orthogonal to the dynamical modes. The free
fields of Sec.~\ref{sec:integral} are understood to contain no such static component, and
over the dispersion branches
that remain the expansion is complete for them.

Orthogonality removes the cross terms from both integrals, but it does not by itself reduce
the surviving terms to the cycle averages~\r{avgs}: each of them carries a weight. For the energy
that weight is the squared modal amplitude $|a_n(\_k)|^2$ written out in Eq.~\r{totalW}.
For the flux there is a second factor, because $U$ is obtained by transforming in $x$, $y$
and $t$ rather than in the three spatial coordinates. At each $(\o,k_x,k_y)$ the modes that
contribute are the roots $k_z$ of $\o_n(\_k)=\o$, and restoring the integration over $k_z$
in place of the one over $\o$ costs the Jacobian of that change of variables,
$\d\o/\d k_z=\d\o_n/\d k_z$, so that
\e
U=\frac14\sum_n\int\frac{{\rm d}^3k}{(2\pi)^3}
\frac{\left|a_n(\_k)\right|^2\,\_\Psi_n^{\dagger}\mathsf L_z\_\Psi_n}
{\left|\d\o_n/\d k_z\right|} ,
\l{totalU}
\f
the counterpart of Eq.~\r{totalW}. Each term is the cycle-averaged flux $\avg{S_z}$ of
Eq.~\r{avgs} for that mode, given by Eq.~\r{AGflux}, divided by $|\d\o_n/\d k_z|$; and since
$\avg{S_z}/\avg{W}=\d\o_n/\d k_z=v_{g,z}^{(n)}$ by Eq.~\r{vg}, each term equals
\e
\frac{\left|a_n\right|^2\_\Psi_n^{\dagger}\mathsf L_z\_\Psi_n}{4\left|\d\o_n/\d k_z\right|}
={\rm sgn}\!\left(v_{g,z}^{(n)}\right)
\frac{\left|a_n\right|^2\_\Psi_n^{\dagger}\mathsf M\_\Psi_n}{4} .
\l{Usign}
\f
Comparison with Eq.~\r{totalW} then shows that the energy carried across the plane is the energy
of the components travelling towards $+z$ minus the energy of those travelling towards
$-z$, which is what the indefiniteness of $\mathsf L_z$ expresses.

\section{Numerical parameters}
\label{app:hydronum}

The code that produces Figs.~\ref{fig:hydroA} and~\ref{fig:hydroB} uses normalized units in
which $\E_0=\M_0=1$ and
$\o_p=1$, so that $c=1$; frequencies and wave numbers are given in units of $\o_p$ and
$k_p=\o_p/\beta$, and the driving field is normalized to $|E_z|=1$. The ratio $\beta/c$
enters Fig.~\ref{fig:hydroA}(a) alone, where it is taken to be $2.8\x10^{-3}$, which is the
value of
$\beta=\sqrt{3/5}\,v_{\rm F}$ for sodium. Every other panel measures wave numbers in units of
$k_p$, and none of the comparisons shown in them depends on the value of $\beta$.

Figure~\ref{fig:hydroA}(a) is obtained directly from the two branches $\o^2=\o_p^2+c^2k^2$ and
$\o^2=\o_p^2+\beta^2k^2$ of the lossless medium, with $ck/\o_p$ swept over
$10^{-2}\le ck/\o_p\le2.5\x10^3$ on a logarithmic axis.

Figure~\ref{fig:hydroA}(b) is computed at the fixed real frequency $\o=1.2\,\o_p$ with $k$
swept over
$0<k\le3\,k_p$, so that the constitutive pole $\o=\beta k$ falls at $k=1.2\,k_p$, for the
collision rates $\nu/\o_p=0.02$ and $0.1$; the heating rate is plotted in units of
$\E_0\o_p|E_z|^2$. Figure~\ref{fig:hydroA}(c) is computed at $\nu/\o_p=0.1$ with $\o$ swept
over
$1.002\,\o_p\le\o\le2.6\,\o_p$, the complex root $k$ of Eq.~\r{hydrocomplexk} being selected
by $\Im k>0$. The two collision rates have been chosen so that the curves separate visibly;
in metals, $\nu/\o_p$ lies nearer to the lower end of this range.

The three panels of Fig.~\ref{fig:hydroB} follow the lossless longitudinal branch, $\nu=0$,
with $k$ swept over the
same interval $0<k\le3\,k_p$. For each $k$ the mode fields are constructed from
$\_J=\ii\o\E_0\_E$ and $\varrho=\ii k\E_0E_z$, the three contributions to Eq.~\r{hydroW} are
normalized to their sum in panel (a), the flux is plotted in units of $\E_0\beta|E_z|^2/2$ in
panel (b), and the velocity in units of $\beta$ in panel (c). The derivatives
$\d(\o\E_{\rm L})/\d\o$ and $\d\E_{\rm L}/\d k$ are evaluated by central differences, so
that the AG side of each comparison is computed from the function $\dyad\E(\o,\_k)$ as a
black box.

The hydrodynamic description of the electron fluid presumes that the phase velocity of the longitudinal wave exceeds the Fermi velocity
$v_{\rm F}$, since otherwise the wave is strongly Landau damped by single-particle
excitations, which the model does not contain. On the longitudinal branch
$\o/k=\beta\sqrt{1+(k_p/k)^2}$, so that with $\beta=\sqrt{3/5}\,v_{\rm F}$ this condition
reads $k\lesssim1.2\,k_p$. For sodium one has, in addition, $k_p\simeq1.2\,k_{\rm F}$, so
that the whole of the range shown is comparable with, or larger than, the Fermi wave number.
The large-$k$ parts of Figs.~\ref{fig:hydroA}(b), \ref{fig:hydroA}(c) and~\ref{fig:hydroB}
should therefore be read
as an illustration of the identities within the model rather than as a prediction for a
specific metal.

\bibliography{refs}

\end{document}